\documentclass[11pt]{article}
\usepackage[preprint]{acl}
\usepackage{times}
\usepackage{latexsym}
\usepackage[T1]{fontenc}
\usepackage[utf8]{inputenc}
\usepackage{microtype}
\usepackage{inconsolata}
\usepackage{graphicx}
\usepackage{amsmath}
\usepackage{amssymb}
\usepackage{booktabs}
\usepackage{array}
\usepackage{colortbl}
\usepackage{tikz}

\usetikzlibrary{patterns}
\definecolor{TableInk}{HTML}{202020}
\definecolor{TableMuted}{HTML}{526675}
\definecolor{TableRule}{HTML}{404040}
\definecolor{TableStage}{HTML}{EAF1F8}
\definecolor{TableOpening}{HTML}{E8F3EB}
\definecolor{TableDuring}{HTML}{F0EBF8}
\definecolor{TableClosing}{HTML}{F9ECEF}
\definecolor{TableBoundary}{HTML}{EAF1F8}
\definecolor{TableJev}{HTML}{B74F18}
\definecolor{TableAdopted}{HTML}{356B4C}
\definecolor{TableRepair}{HTML}{44799D}
\definecolor{TableJudging}{HTML}{9273BA}
\definecolor{TableNegative}{HTML}{A95164}
\definecolor{TableTrack}{HTML}{E7EBEE}
\definecolor{TableNeutral}{HTML}{9AA5AE}
\definecolor{TableBarUp}{HTML}{7FB69A}
\definecolor{TableBarDown}{HTML}{D28A98}

\newcommand{\TableGroupSeparatorSingle}{%
\addlinespace[2pt]\specialrule{0.4pt}{0pt}{2pt}}

\newcommand{\SignedBar}[3]{%
\begin{tikzpicture}[x=1pt,y=1pt,baseline=-2pt]
\path[use as bounding box] (-32,-2) rectangle (32,5);
\draw[TableTrack,line width=0.6pt] (-32,1.5)--(32,1.5);
\fill[#3] (0,-0.5) rectangle (#2,3.5);
\draw[TableRule,line width=0.75pt] (0,-2)--(0,5);
\end{tikzpicture}}
\newcommand{\SignedAxis}[2]{%
\makebox[64pt]{\makebox[26pt][l]{#1}\makebox[12pt][c]{0}\makebox[26pt][r]{#2}}}

\newlength{\TableJudgingNumberWidth}
\newcommand{\JudgingDagger}{%
\begingroup\settowidth{\TableJudgingNumberWidth}{0.000}%
\makebox[\TableJudgingNumberWidth][c]{$\dagger$}\endgroup}

\title{JevCoder: When Does Decision Delegation Pay Off in Code Repair?}
\author{Jiahao Zhang \qquad Yifan Zhang \qquad Yu Huang \\
  {\normalfont Vanderbilt University, Nashville, Tennessee, USA} \\
  \texttt{\{jiahao.zhang,yifan.zhang.2,yu.huang\}@vanderbilt.edu}}
\hypersetup{
  pdftitle={JevCoder: When Does Decision Delegation Pay Off in Code Repair?},
  pdfauthor={Jiahao Zhang, Yifan Zhang, Yu Huang}
}
\begin{document}
\maketitle
\begin{abstract}
A coding agent repeatedly makes decisions as it repairs code. These
decisions shape repair trajectories and affect the token cost of successive model
requests. We ask which repair decisions can be delegated to a low-cost probabilistic
judge to reduce token use while maintaining repair quality. The judges answer questions with
probabilities over predefined answers, without generating explanations.
We evaluate judgment quality across repair stages and test how individual
interventions change subsequent repair work. We compare these interventions with unguided repair, which uses no judge,
and with the same workflow using the repair model itself as the judge. The
benefit depends on the work an intervention changes, rather than the cost of
the judgment alone. Based on these studies, we propose JevCoder, a workflow
built around Jev, a low-cost probabilistic judge. It combines code-snippet
recommendations with action-type guidance and uses judgments to determine when
to submit. We evaluate three repair models, each on the same 400 tasks: 200
from SWE-bench Verified and 200 from SWE-Bench Pro Verified. Compared to unguided repair, JevCoder reduces mean repair-model token use
by 21.6\%--50.6\%, with higher observed repair rates across all three models.
\end{abstract}

\section{Introduction}
\label{sec:intro}

\begin{figure}[!t]
  \centering
  \includegraphics[trim=0 5bp 0 5bp,clip,width=75mm]{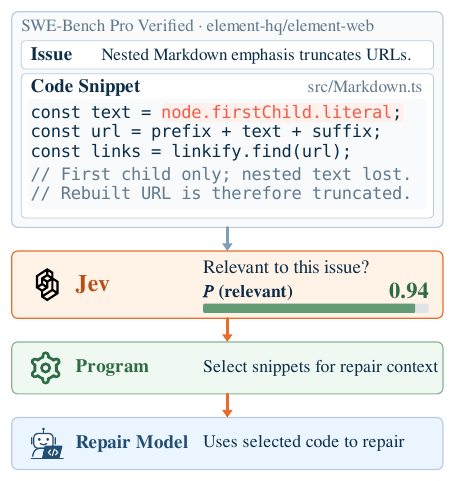}
  \caption{In a simplified recorded SWE-Bench Pro Verified case,
  JevCoder uses Jev's relevance probabilities to rank candidate snippets
  and supply selected code as the coding agent's initial context.}
  \label{fig:recorded-snippet-judgment}
\end{figure}

\begin{figure*}[!t]
  \centering
  \includegraphics[width=\textwidth]{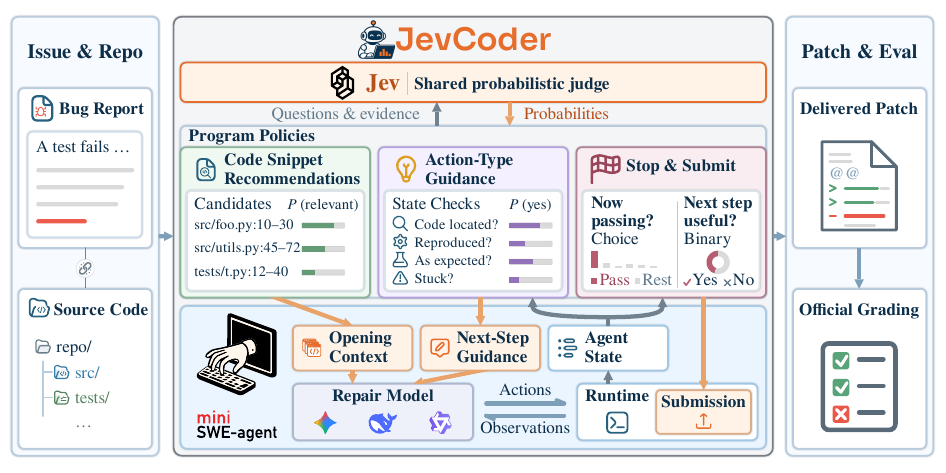}
  \caption{JevCoder delegates judgments to a shared probabilistic judge:
  Jev. Before repair, a program retrieves candidate code for the issue,
  Jev scores each snippet's relevance, and the highest-scoring snippets
  enter the repair model's initial context. After each command, program rules select relevant state questions for
  Jev and use its answers to suggest an action type. The repair model
  chooses the specific command. The stopping policy depends on the model class. For the larger API models (DeepSeek, Gemini),
  a Choice question on the latest check output (Pass: a previously
  failing check now passes; Rest: other outcomes) triggers direct
  submission. For the smaller model (Qwen), repeated No answers to whether the next step adds
  information trigger a reminder and then submission.}
  \label{fig:jevcoder-overview}
\end{figure*}

A coding agent's repair model chooses where to look, what to try, and
when to stop. These choices affect how many steps the agent takes.
When workflows resend interaction history, each request processes
the accumulated context.
Workflow design substantially affects this inference cost
\citep{agentless2024,codemonkeys}. Assigning selected judgments
to a cheaper model could reduce this expense. Whether delegation pays off depends on whether repair quality is
maintained and on how it changes subsequent repair work.

Coding agents receive feedback during repair or use verifiers to select
candidate patches \citep{swe_gym,r2e_gym}. SWE-PRM finds limited benefits
from same-model supervision, but consistent gains with a stronger
supervisor. In one setting, supervisor calls cost over six times as much as
repair-model calls \citep{swe_prm}.
We therefore ask which repair decisions are worth delegating to a low-cost judge.

We study judges that return probabilities over predefined answers
without generating explanations.
Jev, TypeSafe's decision model, provides structured answers and probabilities \citep{typesafe2026}.
For binary questions, Jev returns the probability of ``yes''.
Choice returns probabilities for predefined options. Score uses a
predefined rating scale, with a probability assigned to each level. Our experiments evaluate
all three forms. JevCoder uses binary and Choice judgments to rank
snippets or trigger repair interventions.
We also use the repair model itself as a judge, reading probabilities
for predefined answer tokens.
Program rules determine when to ask a question and how to use the answer.
Figure~\ref{fig:recorded-snippet-judgment} illustrates this process for
snippet recommendations.
For the URL truncation issue, Jev assigns relevance 0.94 to the snippet
that reads only the first child of an emphasis node.

We assess judgment quality across repair stages and examine how individual
interventions change subsequent repair work. Single-component experiments compare
repair outcomes and resource use with unguided repair and, where
available, same-model judging.
They show that more accurate file rankings do not necessarily improve
repair outcomes, and that retaining recommended snippets can reduce
repair-model token use despite increasing context length.
Building on these component results, we propose \textbf{JevCoder} for
cost-efficient code repair (Figure~\ref{fig:jevcoder-overview}).
It uses code-snippet recommendations to guide
initial code exploration and provides action-type guidance during repair.
Its stopping policies use judgments of successful checks or repeated
lack of progress to determine when the agent should end a repair attempt
and submit its current patch.

For each of three repair models, we compare JevCoder with unguided
repair on the same 400 tasks: 200 from SWE-bench Verified and 200 from
SWE-Bench Pro Verified. JevCoder reduces mean repair-model token use by
21.6\%--50.6\% and achieves higher observed repair rates for all three
models. Compared with same-model judging, which helps distinguish
workflow benefits from judging overhead, JevCoder uses 9.1\% fewer tokens
with DeepSeek and 23.3\% fewer with Qwen.
Code and evaluation results are available at
\url{https://anonymous.4open.science/r/jevcoder-artifact-8E62}.

\section{Related Work}
\label{sec:related}

\paragraph{Judgment in coding agents.}
Coding agents use feedback during repair and verifiers to select completed
patches.
SWE-PRM reads the issue and recent trajectory to diagnose errors,
then returns natural-language feedback for the repair model's next steps
\citep{swe_prm}. Trained small critics can also provide this guidance
\citep{steer2026}.
For completed candidates, SWE-Gym trains a verifier on successful and
failed repair trajectories. Given the issue, a completed trajectory and
its diff, the verifier uses answer probabilities to score candidate
repairs \citep{swe_gym}.
Testing adds execution evidence: reproduction tests for the issue
and regression tests for existing behavior \citep{agentless2024}.
Hybrid verification combines this
evidence with learned scores to select a candidate \citep{r2e_gym}.
Replacing per-step model decisions with cheaper mechanisms is also
studied outside repair: AutoTool predicts the next tool from transition
statistics of past trajectories with minimal model inference
\citep{autotool2026}, and query routing sends requests to cheaper
models when they suffice \citep{frugalgpt2024,routellm}.
JevCoder uses answer probabilities from fixed questions to trigger predefined
interventions, which we evaluate across repair stages.

\paragraph{LLM judgment and probability reliability.}
LLM judges evaluate responses through scores or pairwise comparisons
\citep{mtbench2023}. Scores can be computed from probabilities assigned
to rating tokens or predefined answers \citep{geval2023,swe_gym}.
In instruction-tuned models, the answer token with the highest probability
can differ from the answer expressed in generated text
\citep{myanswer2024}.
For software repair, SWE-RM separately evaluates how well scores
distinguish successful from failed repairs and how closely predicted
probabilities match observed success rates \citep{swe_rm2026}.
We test judgment quality on specific repair questions and how using
these judgments affects repair outcomes and resource use.

\section{Study Setting}
\label{sec:setup}
\paragraph{Tasks and comparisons.}
We evaluate decision delegation on repository-level issue repair using
a fixed sample of 200 tasks from SWE-bench Verified
\citep{swebenchverified2024} and 200 from SWE-Bench Pro Verified
\citep{zheng2026swebenchproverified}. Our sample covers Python, Go, JavaScript, and TypeScript repositories. The repair models are
DeepSeek-V4.1-Flash \citep{deepseekv412026}, Qwen3.8-27B-FP8
\citep{qwen3827b2026}, and Gemini~3.8 Flash \citep{gemini38flash2026}.
For each model, full-workflow evaluation compares JevCoder with unguided
repair on the same 400 tasks. For DeepSeek and Qwen, we also replace Jev with a same-model judge.
Gemini's API does not expose the answer-token probabilities required for
this comparison.
DeepSeek's component studies use balanced samples of 72 SWE-bench Verified tasks
and 118 SWE-Bench Pro Verified tasks, comprising all failures from the original
unguided runs and an equal number of randomly sampled successes.
Including both groups lets us measure repair gains and losses. Qwen component experiments follow
similar procedures with model-specific samples; Gemini analyses focus
on selected judgments in existing repair trajectories
(Appendix~\ref{app:gemini-stopping-replay}).
The patch-outcome study uses 1,192 submitted-patch records from 343 tasks
in SWE-bench Verified and SWE-Bench Pro v2 \citep{swebenchpro2025}.

\paragraph{Judges and evidence.}
Jev and the same-model judge receive the same question and evidence.
We read the same-model judge's probabilities for predefined answer
tokens. Judgment-quality experiments also include the small LLM
Qwen3.5-2B \citep{qwen352b2026}. Evidence includes, for example,
source snippets for relevance and command history for repetition.
Reference patches and official test outcomes supply evaluation labels. The evidence supplied to each judge and its probability readouts are
described in Appendices~\ref{app:configuration}, \ref{app:protocol},
and~\ref{app:offline}.

\paragraph{Repair and resource measures.}
Repair rate is official pass@1: the fraction of tasks whose delivered
patch passes the benchmark tests. Repair-model tokens count the repair
model's total input and output usage.
Judge tokens are added for the same-model judging condition.
Runtime estimates combine repair-model and tool-execution time with
judge inference time. Remote API timings are adjusted by subtracting measured fixed
connection delays; judge time is reported separately and included in
the full-task total. Estimation and accounting details appear in
Appendices~\ref{app:judge-overhead} and~\ref{app:protocol}.
Each attempt is limited to
250 repair-model requests, a 3,000-second time limit,
and US\$3 in API costs. The first limit reached stops
execution.

\paragraph{Analysis.}
Differences are paired by task, with task-level percentile bootstrap
95\% intervals. Primary summaries are mean token use and the median
of paired runtime differences. Means summarize aggregate token use;
runtime medians describe a typical task-level change.
SWE-Bench Pro Verified opening comparisons are reweighted to the
baseline failure and success proportions in the original 200 tasks.
The 72-task estimates are unweighted. Raw repair counts
refer to the tested sample. Each comparison uses one run per task and
condition. The repair-loss
criterion requires the lower bound of the paired 95\% interval for the
repair-rate difference to be at least $-10$ percentage points. Appendix~\ref{app:component-estimation} provides sampling and analysis
details, including the weights.

\section{Judgment Quality}
\label{sec:judgment-quality}
We first examine which judgments Jev can make reliably at each stage of
repair. The results guide the intervention studies in
Section~\ref{sec:components} and the design of JevCoder in
Section~\ref{sec:jevcoder}.

\begin{table*}[t]
\centering
\begingroup
\fontsize{10}{12}\selectfont
\setlength{\tabcolsep}{3pt}
\renewcommand{\arraystretch}{1.02}
\arrayrulecolor{TableRule}
\begin{tabular}{@{}>{\raggedright\arraybackslash}p{119pt}>{\raggedright\arraybackslash}p{90pt}>{\raggedleft\arraybackslash}p{31pt}>{\raggedleft\arraybackslash}p{54pt}>{\raggedleft\arraybackslash}p{38pt}>{\raggedright\arraybackslash}p{\dimexpr\textwidth-332pt-30pt\relax}@{}}
\toprule[0.8pt]
\textbf{Judgment} & \textbf{Metric} & \textbf{Jev} & \textbf{Comparator} & \textbf{$\Delta$} & \mbox{\textbf{Comparator model}} \\
\midrule
\multicolumn{6}{@{}l@{}}{\rule[-2pt]{0pt}{13pt}\textcolor{TableInk}{\textbf{Opening --- Initial context}}} \\
\addlinespace[1pt]
File relevance (Binary) & Top-1 hit (\%) $\uparrow$ & \textbf{75.9} & 29.0 & $+46.9$ & Qwen3.5-2B \\
\addlinespace[0pt]
File relevance (Score) & Top-1 hit (\%) $\uparrow$ & \textbf{94.9} & 88.1 & $+6.8$ & Jev (Choice)\textsuperscript{\fontsize{8}{9}\selectfont $\ddagger$} \\
\addlinespace[0pt]
Snippet recommendations & Selection recall (\%) $\uparrow$ & \textbf{81.6} & 78.0 & $+3.6$ & DeepSeek \\
\addlinespace[0pt]
\TableGroupSeparatorSingle
\multicolumn{6}{@{}l@{}}{\rule[-2pt]{0pt}{13pt}\textcolor{TableInk}{\textbf{During repair --- State and action judgments}}} \\
\addlinespace[1pt]
Code located & AUROC $\uparrow$ & \textbf{0.738} & 0.701 & $+0.038$ & DeepSeek/Qwen \\
\addlinespace[0pt]
Issue reproduced & AUROC $\uparrow$ & \textbf{0.822} & 0.804 & $+0.018$ & DeepSeek/Qwen \\
\addlinespace[0pt]
Stalled repair & AUROC $\uparrow$ & \textbf{0.968} & 0.931 & $+0.037$ & DeepSeek/Qwen \\
\addlinespace[0pt]
Post-edit check & AUROC $\uparrow$ & \textbf{0.802} & 0.797 & $+0.005$ & DeepSeek/Qwen \\
\addlinespace[0pt]
\TableGroupSeparatorSingle
\multicolumn{6}{@{}l@{}}{\rule[-2pt]{0pt}{13pt}\textcolor{TableInk}{\textbf{Closing --- Submission judgments}}} \\
\addlinespace[1pt]
Patch outcome (Binary) & AUROC $\uparrow$ & 0.595 & \textbf{0.684} & $-0.090$ & DeepSeek \\
\addlinespace[0pt]
Patch outcome (Score) & AUROC $\uparrow$ & 0.642 & \textbf{0.652} & $-0.011$ & DeepSeek \\
\addlinespace[0pt]
Successful-check signal & AUROC $\uparrow$ & \textbf{0.990} & 0.974 & $+0.016$ & DeepSeek \\
\addlinespace[0pt]
\bottomrule[0.8pt]
\end{tabular}
\arrayrulecolor{black}
\endgroup
\caption{Judgment quality and snippet selection.
Higher is better; bold marks the better point estimate.
$\Delta$ is Jev minus the comparator (percentage points for percentage
metrics; absolute differences for AUROC), computed before rounding.
File relevance uses 100 candidate files for Binary and 10 for Score.
Selection recall is measured over reference lines covered by candidate snippets.
State AUROCs pool DeepSeek and Qwen inputs, each evaluated against its same-model judge.
\textsuperscript{\fontsize{8}{9}\selectfont $\ddagger$}\,Jev (Choice) returns probabilities over the same ten candidate files.}
\label{tab:judgment-evidence}

\end{table*}

\subsection{Code Relevance}
Preparing the agent's initial context requires choosing which retrieved
code is relevant to the issue. A judge can support this choice by identifying useful starting points
for the agent's investigation. We first examine this ability at file level,
using files modified by the reference patch as evaluation labels.

We compare Jev with Qwen3.5-2B \citep{qwen352b2026}, a small language
model with similarly low inference overhead. A program retrieves 100
candidate files by combining BM25 retrieval with path and symbol
matching. Both judges answer the same binary relevance question for each
file, which we rank by the probability of a YES answer. On 162
SWE-Bench Pro Verified tasks, the top-ranked file is modified by the
reference patch in 75.9\% of tasks for Jev and 29.0\% for Qwen3.5-2B.
Median judgment time is below 0.15 seconds per file for both judges
(Appendix~\ref{app:judge-overhead}).

File rankings identify where to look. We then assess relevance at a
finer granularity to select individual code snippets for the agent's
initial context. Selection recall is measured
against the source lines that the reference patch removes or replaces;
189 of the 200 SWE-Bench Pro Verified tasks have such lines. Jev's
selected snippets retain 81.6\% of the reference lines covered by the
candidates, compared with 78.0\% for DeepSeek given the same candidates
and questions (Appendix~\ref{app:snippet-coverage}).

\subsection{Repair State and Next-Action Judgments}
What to try next depends on the stage reached during repair.
For action-type guidance, we ask whether the agent has opened code
relevant to the issue (code located) and whether a test or probe shows
the reported failure (issue reproduced). We also assess whether recent
actions repeat without new information (stalled repair) and whether a
check after the latest edit shows the expected behavior (post-edit check).
We derive evaluation labels from reference patches and recorded execution
outcomes. We label code location using successful reads of reference-patch
files, and reproduction using failed target tests or failed probes with
exceptions. For stalled repair, we use command repetition without file
writes; for post-edit checks, we use target checks passing since the
latest edit. Full labeling rules appear in Appendix~\ref{app:offline}.
Each judge's probability of a positive answer is the ranking score.
The area under the receiver operating characteristic curve (AUROC)
measures how often positive states score above negative states, with ties
counting as one half (0.5 indicates random ranking).

Given the same questions and evidence, Jev achieves higher pooled AUROC
than same-model judging for all four state questions across DeepSeek and
Qwen trajectories (Table~\ref{tab:judgment-evidence}). We find the clearest discrimination for
stalled repair (AUROC 0.968), compared with 0.738 for code location.

We also test recognition of repeated commands.
On 883 command pairs derived from recorded repair trajectories, Jev
achieves 99.4\% recall with a 0.6\% false-positive rate at a probability
threshold of 0.7. At the same threshold, Qwen3.5-2B achieves 95.5\% recall with
a 17.6\% false-positive rate (Appendix~\ref{app:judgment-labels}).
Because the positives are rewritten rather than identical commands,
this test covers repetition that an exact-match rule cannot detect.
Before execution, the relevant question is whether a proposed command
adds information or advances repair beyond recent commands and outputs.
We compare judges through replay on 14,170 common command states,
calibrating each threshold to block at most 1\% of non-repeated
commands that write source files. Under this constraint, the replayed policy blocks 61.9\% of the scored steps
in runs containing at least ten consecutive identical commands using Jev's
probabilities, compared with 28.0\% using Qwen3.5-2B's.

\begin{table*}[t]
\centering
\begingroup
\fontsize{10}{12}\selectfont
\setlength{\tabcolsep}{3pt}
\renewcommand{\arraystretch}{1.05}
\arrayrulecolor{TableRule}
\begin{tabular}{@{}>{\raggedright\arraybackslash}m{130pt}>{\raggedright\arraybackslash}m{48pt}>{\centering\arraybackslash}m{18pt}>{\centering\arraybackslash}m{71pt}>{\centering\arraybackslash}m{95pt}>{\centering\arraybackslash}m{\dimexpr\textwidth-362pt-30pt\relax}@{}}
\toprule[0.8pt]
\textbf{Intervention} & \textbf{Model} & $\boldsymbol{N}$ & \textbf{$\Delta$Repair} (pp) $\uparrow$ & \textbf{$\Delta$Tokens} (M) $\downarrow$ & \textbf{In JevCoder} \\
 & & & \SignedAxis{$-10$}{$+10$} & \SignedAxis{$-1.5$}{$+1.5$} & \\
\midrule
\multicolumn{6}{@{}l@{}}{\rule[-2pt]{0pt}{13pt}\textcolor{TableInk}{\textbf{Opening --- Initial context}}} \\
\addlinespace[0.5pt]
File recommendations (Binary) & DeepSeek & 118 & \shortstack{$+0.50$\\[-4pt]\SignedBar{0.5}{1.600000}{TableNeutral}} & \shortstack{$-0.354$\\[-4pt]\SignedBar{-0.354}{-7.552000}{TableBarUp}} & \textcolor{TableMuted}{---} \\[-1pt]
\addlinespace[0pt]
File recommendations (Score) & DeepSeek & 118 & \shortstack{$-3.58$\\[-4pt]\SignedBar{-3.58}{-11.456000}{TableBarDown}} & \shortstack{$-0.372$\\[-4pt]\SignedBar{-0.372}{-7.936000}{TableBarUp}} & \textcolor{TableMuted}{---} \\[-1pt]
\addlinespace[0pt]
Snippet recommendations & DeepSeek & 118 & \shortstack{$+3.89$\\[-4pt]\SignedBar{3.89}{12.448000}{TableBarUp}} & \shortstack{$-0.206$\\[-4pt]\SignedBar{-0.206}{-4.394667}{TableBarUp}} & \textcolor{TableAdopted}{{\fontsize{8.5}{10}\selectfont$\checkmark$}} \\[-1pt]
\addlinespace[0pt]
\TableGroupSeparatorSingle
\multicolumn{6}{@{}l@{}}{\rule[-2pt]{0pt}{13pt}\textcolor{TableInk}{\textbf{During repair --- Next-action interventions}}} \\
\addlinespace[0.5pt]
Command blocking & Qwen & 72 & \shortstack{$-2.78$\\[-4pt]\SignedBar{-2.78}{-8.896000}{TableBarDown}} & \shortstack{$-0.195$\\[-4pt]\SignedBar{-0.195}{-4.160000}{TableBarUp}} & \textcolor{TableMuted}{---} \\[-1pt]
\addlinespace[0pt]
Action-type guidance & DeepSeek & 72 & \shortstack{$+8.33$\\[-4pt]\SignedBar{8.33}{26.656000}{TableBarUp}} & \shortstack{$-0.169$\\[-4pt]\SignedBar{-0.169}{-3.605333}{TableBarUp}} & \textcolor{TableAdopted}{{\fontsize{8.5}{10}\selectfont$\checkmark$}} \\[-1pt]
\addlinespace[0pt]
\TableGroupSeparatorSingle
\multicolumn{6}{@{}l@{}}{\rule[-2pt]{0pt}{13pt}\textcolor{TableInk}{\textbf{Closing --- Submission policies}}} \\
\addlinespace[0.5pt]
Submission reminder & DeepSeek & 72 & \shortstack{$+6.94$\\[-4pt]\SignedBar{6.94}{22.208000}{TableBarUp}} & \shortstack{$+0.001$\\[-4pt]\SignedBar{0.001}{0.021333}{TableNeutral}} & \textcolor{TableMuted}{---} \\[-1pt]
\addlinespace[0pt]
Automatic submission & DeepSeek & 72 & \shortstack{$+8.33$\\[-4pt]\SignedBar{8.33}{26.656000}{TableBarUp}} & \shortstack{$-0.295$\\[-4pt]\SignedBar{-0.295}{-6.293333}{TableBarUp}} & \textcolor{TableAdopted}{{\fontsize{8.5}{10}\selectfont$\checkmark$}} \\[-1pt]
\addlinespace[0pt]
No-progress submission & Qwen & 72 & \shortstack{$+0.00$\\[-4pt]\SignedBar{0.0}{0.000000}{TableNeutral}} & \shortstack{$-1.258$\\[-4pt]\SignedBar{-1.258}{-26.837333}{TableBarUp}} & \textcolor{TableAdopted}{{\fontsize{8.5}{10}\selectfont$\checkmark$}} \\[-1pt]
\addlinespace[0pt]
\bottomrule[0.8pt]
\end{tabular}
\arrayrulecolor{black}
\endgroup
\caption{Component comparisons relative to matched unguided repair on $N$ tasks.
Values are changes in repair rate (pp) and in mean repair-model tokens
(M); bars show the direction and size of each change. Green and red
mark improvement and degradation; grey marks small changes (absolute
change below 1~pp or 0.01M). A check mark denotes adoption in JevCoder.
95\% intervals and full comparisons appear in Appendix~\ref{app:component-selection}.}
\label{tab:component-design-evidence}

\end{table*}

\subsection{Stopping and Submission Judgments}
Deciding when to submit raises a broader question: will the current patch
resolve the issue? We compare predicted probabilities of passing official
tests with official repair outcomes on the patch sample described in
Section~\ref{sec:setup}. Across four input configurations, Jev shows limited discrimination,
with an AUROC of 0.562--0.595, compared with 0.638--0.684 for DeepSeek
(Appendix~\ref{app:patch-outcomes}).

A finer answer format could recover part of the gap. Jev's Score
interface returns probabilities over ordered levels of test success. On 840 common DeepSeek patch records, we use the probability of all target
tests passing as the ranking score. Jev achieves an AUROC of 0.642, up from
0.581 for its binary question on the same records and close to that of DeepSeek scoring.
The ordered question improves discrimination, but Jev still struggles to
predict whether a patch will pass the official tests.

We therefore examine whether Jev can recognize that a previously
failing check now passes.
On 52 matched states from DeepSeek trajectories, Jev distinguishes
these transitions with an AUROC of 0.990, compared with 0.974 for
DeepSeek judging (Table~\ref{tab:judgment-evidence};
Appendix~\ref{app:successful-check-quality}).
Jev is near-perfect when the answer is recorded in the
history and weak when it must be inferred.

\section{Decision Delegation Across Repair Stages}
\label{sec:delegation}
\label{sec:components}
Judgment quality alone does not show whether delegation reduces the work
needed for repair. We therefore evaluate each intervention separately
against unguided repair, following repair from initial code exploration
to submission. Each comparison links the change in repair work to token
use and repair outcomes.
Complete component comparisons and the evidence for selecting these
components appear in Appendix~\ref{app:component-selection}.

\subsection{Code-Snippet Recommendations}
Useful code recommendations should reduce the repository exploration
needed for repair. We first rank 100 candidate files by Jev's relevance
probabilities and supply retrieved source regions from the five highest-ranked
files. With DeepSeek as the repair model, file recommendations use
5.8 fewer repair-model requests and 0.354M fewer
repair-model tokens per task on average, while resolving a similar number of issues.
File recommendations using Jev's Score interface also reduce token use,
but do not meet the prespecified repair-loss criterion
(Table~\ref{tab:component-design-evidence}).
Once the right files are open, the remaining decisions concern what to
change inside them, which a file ranking does not inform
(Appendix~\ref{app:file-scoring}).

We therefore evaluate
recommendations that select individual code snippets from retrieved source regions using Jev's relevance probabilities. On the same sample, snippet
recommendations resolve 68 issues, compared with 64 for file
recommendations and 63 without recommendations. Relative to unguided
repair, the weighted repair rate is 3.9 percentage points higher.
Snippet recommendations also use
0.206M fewer repair-model tokens per task on average, with comparable
runtime (Appendix~\ref{app:opening-evidence}).

Recommended snippets stay in the interaction history, so every later
request carries their tokens. We therefore examine whether they reduce
subsequent investigation by more than they add.
Snippet recommendations also reduce read and search commands before the first editing command by 5.0 per task on average,
with 0.126M fewer repair-model tokens over this period.
On 116 matched DeepSeek tasks, removing the recommended snippets after ten
repair-model requests increases mean repair-model token use by 0.538M per task
(95\% CI: 0.188M--0.930M), compared with retaining them throughout repair,
and increases the paired median runtime by 32.1 seconds.
Withdrawing the snippets increases subsequent investigation, so keeping
them available reduces overall token use.
On the 118-task Qwen comparison, mean repair-model token use is
0.230M lower per task with no loss in repair rate
(Appendix~\ref{app:qwen-opening}).

\subsection{Action-Type Guidance}
Initial code provides a starting point, but the next useful action
depends on what the agent learns during repair.
Command blocking prevents execution of a command judged unlikely to
add information and asks the model to propose an alternative. On Qwen,
the lower bound of the repair-rate interval for this intervention falls
below $-10$ percentage points
(Table~\ref{tab:component-design-evidence};
Appendix~\ref{app:repetition-blocking}).
Jev returns only a probability, so blocking can reject a command but
cannot say what to do instead.
After a tool result,
Jev judges the current repair state, and a program maps its answers to
action-type suggestions, such as reproducing the issue or editing code.
The repair model chooses the specific command.

With DeepSeek, Jev guidance resolves
45 issues, compared with 39 for unguided repair and 44 for same-model judging. Relative to unguided repair, the repair rate is 8.3 percentage
points higher, and it uses 0.169M fewer repair-model
tokens per task on average. Same-model judging adds 18.4 seconds to the paired median runtime
relative to unguided repair; relative to same-model judging, Jev
guidance shortens it by 26.1 seconds
(Appendix~\ref{app:action-guidance}).

\subsection{Submission and Early Stopping}
\label{sec:ending}
We next examine whether timely submission can avoid unnecessary
repair-model requests. A previously failing test or reproduction check
that now passes signals that repair may be complete. The agent must
still decide to stop and submit. In unguided runs, Jev's
successful-check criterion is met in 130 of 400 DeepSeek runs, 57 of
400 Qwen runs, and 156 of 400 Gemini runs. After the first such check,
the repair model still makes a median of nine, five, and 14 further
requests, respectively.

When Jev's successful-check judgment meets the submission rule, the
program can remind the repair model to submit or submit directly. With DeepSeek as the repair model, half of Jev's 32 reminders are followed by submission within five subsequent
repair-model requests.
The program automatically submits the current patch after a qualifying
judgment, ending repair without another repair-model request.
The workflow resolves 45 issues, compared with 39 for unguided repair,
and uses 0.295M fewer repair-model tokens per task on average (Appendix~\ref{app:submission-details}).

Stalled attempts without a successful-check signal need another basis
for stopping. We test no-progress submission with Qwen3.8-27B, the smaller model in
our study, whose runs more often stall before reaching a passing check. We ask the same question used for command blocking, but use consecutive
low probabilities as a signal to end repair. Repeated no-progress
judgments first trigger a reminder, and
continued lack of progress leads the program to submit the current
patch before executing the proposed command.
No-progress submission and unguided repair each resolve 27 of the 72
SWE-Bench Pro Verified tasks (Appendix~\ref{app:qwen-stopping}). However, it
reduces mean repair-model token use by 1.258M per task (95\% CI:
0.905M--1.633M) and paired median runtime by 655.8 seconds (95\% CI:
79.1--1,866.4).
\section{JevCoder}
\label{sec:jevcoder}
\label{sec:harness}
JevCoder combines the selected components into a complete repair
workflow (Figure~\ref{fig:jevcoder-overview}). Jev supplies probabilities
for fixed questions, and program rules use them to decide how to
intervene. Program rules handle decisions that can be determined directly from the
trajectory. Jev interprets the evidence needed for the remaining judgments. The
repair model generates code and chooses the commands to execute. Jev receives question-specific evidence from the repository
and repair trajectory.
All three repair models use snippet recommendations and action-type
guidance. The stopping policy depends on the class of repair model: the two
larger API models, DeepSeek and Gemini, stop on an observed successful
check, and the smaller open-weight model, Qwen3.8-27B, stops on
repeated lack of progress.

\paragraph{Code-snippet recommendations.}
Before repair begins, a program retrieves source code using the issue
and extracts candidate snippets from the retrieved source regions. Jev
estimates whether each snippet is relevant to the issue. The program selects snippets in descending probability order within a
length budget, merging overlapping source ranges. Selected snippets retain their file paths and line
ranges, so the repair model can inspect the surrounding code. They are
supplied before the first repair-model request and remain in the
interaction history throughout repair.

\paragraph{Action-type guidance.}
After each command, the program selects state questions based on the
repair trajectory. It uses probability thresholds to map Jev's answers
to suggested action types. For example, when repair stalls before an edit, the program suggests
editing if relevant code has been located and the issue reproduced.
If the code is located but the issue has not been reproduced, it suggests
probing instead. The program detects exact command repetition. Jev can identify commands
that repeat searches or reads without obtaining new information, even
when their wording differs. When consecutive judgments indicate stalled repair or a need to switch
action types, the program sends a short note; the repair model chooses
the command.

\ifdefined\WorkflowTableInline
\begin{minipage}{\textwidth}
\captionsetup{type=table}
\else
\begin{table*}[t]
\fi
\centering
\begingroup
\fontsize{10}{12}\selectfont
\setlength{\tabcolsep}{3pt}
\renewcommand{\arraystretch}{1.0}
\arrayrulecolor{TableRule}
\newcommand{\WorkflowChange}[1]{%
\textcolor{black!65}{\fontsize{9}{10.8}\selectfont #1}}
\newcommand{\WorkflowMetric}[4]{%
\makebox[#1][r]{#3}\hspace{2pt}\makebox[#2][l]{\WorkflowChange{#4}}}
\newcommand{\WorkflowNoJudging}{%
\begingroup\settowidth{\TableJudgingNumberWidth}{0.000}%
\makebox[\TableJudgingNumberWidth][c]{---}\endgroup}
\begin{tabular}{@{}>{\raggedright\arraybackslash}m{94pt}>{\centering\arraybackslash}m{72pt}>{\raggedleft\arraybackslash}m{46pt}>{\raggedleft\arraybackslash}m{44pt}>{\centering\arraybackslash}m{88pt}>{\centering\arraybackslash}m{\dimexpr\textwidth-344pt-30pt\relax}@{}}
\toprule[0.8pt]
\textbf{Condition} & \textbf{Repair (\%)} $\uparrow$ & \multicolumn{3}{c}{\textbf{Repair-model tokens (M)} $\downarrow$} & \textbf{Runtime (s)} $\downarrow$ \\
\cmidrule(lr){3-5}
 & & \textcolor{black!80}{Repair} & \textcolor{black!80}{Judging} & \textbf{Total} & \\
\midrule
\multicolumn{6}{@{}l@{}}{\rule[-2pt]{0pt}{13pt}\textcolor{TableInk}{\textit{\textbf{DeepSeek-V4.1-Flash}}}} \\
\addlinespace[0.5pt]
Baseline & \WorkflowMetric{26pt}{40pt}{76.25}{} & 1.978 & \WorkflowNoJudging & \WorkflowMetric{26pt}{48pt}{1.978}{} & \WorkflowMetric{26pt}{48pt}{181.7}{} \\
\addlinespace[0pt]
Same-model judging & \WorkflowMetric{26pt}{40pt}{77.00}{($+0.75$)} & 1.441 & 0.265 & \WorkflowMetric{26pt}{48pt}{1.706}{($-13.7\%$)} & \WorkflowMetric{26pt}{48pt}{287.6}{($+58.2\%$)} \\
\addlinespace[0pt]
\textbf{JevCoder} & \WorkflowMetric{26pt}{40pt}{\textbf{77.25}}{($+1.00$)} & 1.550 & \JudgingDagger & \WorkflowMetric{26pt}{48pt}{\textbf{1.550}}{($-21.6\%$)} & \WorkflowMetric{26pt}{48pt}{\textbf{162.7}}{($-10.5\%$)} \\
\addlinespace[0pt]
\specialrule{0.4pt}{1pt}{2pt}
\multicolumn{6}{@{}l@{}}{\rule[-2pt]{0pt}{13pt}\textcolor{TableInk}{\textit{\textbf{Qwen3.8-27B}}}} \\
\addlinespace[0.5pt]
Baseline & \WorkflowMetric{26pt}{40pt}{45.00}{} & 1.495 & \WorkflowNoJudging & \WorkflowMetric{26pt}{48pt}{1.495}{} & \WorkflowMetric{26pt}{48pt}{965.1}{} \\
\addlinespace[0pt]
Same-model judging & \WorkflowMetric{26pt}{40pt}{45.25}{($+0.25$)} & 0.673 & 0.291 & \WorkflowMetric{26pt}{48pt}{0.964}{($-35.5\%$)} & \WorkflowMetric{26pt}{48pt}{818.4}{($-15.2\%$)} \\
\addlinespace[0pt]
\textbf{JevCoder} & \WorkflowMetric{26pt}{40pt}{\textbf{46.75}}{($+1.75$)} & 0.739 & \JudgingDagger & \WorkflowMetric{26pt}{48pt}{\textbf{0.739}}{($-50.6\%$)} & \WorkflowMetric{26pt}{48pt}{\textbf{641.3}}{($-33.6\%$)} \\
\addlinespace[0pt]
\specialrule{0.4pt}{1pt}{2pt}
\multicolumn{6}{@{}l@{}}{\rule[-2pt]{0pt}{13pt}\textcolor{TableInk}{\textit{\textbf{Gemini 3.8 Flash}}}} \\
\addlinespace[0.5pt]
Baseline & \WorkflowMetric{26pt}{40pt}{68.25}{} & 5.422 & \WorkflowNoJudging & \WorkflowMetric{26pt}{48pt}{5.422}{} & \WorkflowMetric{26pt}{48pt}{490.8}{} \\
\addlinespace[0pt]
\textbf{JevCoder} & \WorkflowMetric{26pt}{40pt}{\textbf{71.00}}{($+2.75$)} & 3.966 & \JudgingDagger & \WorkflowMetric{26pt}{48pt}{\textbf{3.966}}{($-26.9\%$)} & \WorkflowMetric{26pt}{48pt}{\textbf{384.6}}{($-21.6\%$)} \\
\addlinespace[0pt]
\bottomrule[0.8pt]
\end{tabular}
\arrayrulecolor{black}
\endgroup
\caption{JevCoder evaluation on 400 matched tasks per condition.
Tokens are per-task means; runtimes are medians.
Parentheses indicate changes from Baseline
(repair: pp; token means and runtime medians: \%).
Bold marks the best repair rate, Total tokens and runtime within each model.
$\dagger$: Jev tokens are excluded from repair-model totals and reported separately (\S\ref{sec:jev-usage}).}
\label{tab:jevcoder-evaluation}

\ifdefined\WorkflowTableInline
\end{minipage}
\else
\end{table*}
\fi

\paragraph{Submission and early stopping.}
DeepSeek and Gemini use observed check outcomes to end repair. After a
command recognized as a test or reproduction run, Jev answers a
Choice question over outcome categories. If the most probable category
is that a previously
failing check now passes after a source edit, and its probability meets
the submission threshold, the program submits the nonempty current
source patch directly.
The smaller model reaches a successful check in fewer runs and stalls
more often, so Qwen uses a different signal: whether a proposed command will add information
or advance repair beyond recent commands and their outputs. Consecutive
low probabilities first trigger a reminder. A longer sequence triggers
submission of the current patch before executing the proposed command.

\section{JevCoder Evaluation}
\label{sec:jevcoder-evaluation}

The component studies identify interventions that change repair work.
We now examine whether combining them in JevCoder reduces the resources
needed to resolve issues while maintaining repair quality.
Table~\ref{tab:jevcoder-evaluation} compares the full JevCoder workflow with
unguided repair.
Same-model judging with DeepSeek and Qwen helps distinguish workflow
benefits from judging overhead.

\paragraph{Resource use and repair quality.}
JevCoder reduces mean repair-model token use by 21.6\%--50.6\% across
all three models. Qwen shows the largest relative saving, with a
statistically significant 50.6\% reduction. JevCoder saves 0.756M
repair-model tokens per task on average with Qwen
(95\% paired bootstrap CI: 0.590M--0.931M).
The 95\% intervals for DeepSeek and Gemini's mean reductions also
exclude zero (Appendix~\ref{app:full-workflow}). JevCoder uses fewer tokens and achieves higher observed repair rates.
All three JevCoder comparisons meet the prespecified repair-loss criterion.
JevCoder newly resolves more tasks than it loses for every model and
keeps 95.4\% of DeepSeek's and 94.9\% of Gemini's baseline-resolved
tasks (Appendix~\ref{app:task-outcomes}).
Runtime includes tool execution and judge inference. JevCoder reduces paired median
runtime by 149.3 seconds for Qwen, 123.2 seconds for Gemini, and
9.5 seconds for DeepSeek; all three intervals exclude zero
(Appendix~\ref{app:full-workflow}).

\paragraph{Same-model judging.}
Same-model judging tests whether the workflow can reduce token use when
the repair model answers the same fixed questions. Relative to unguided repair, these conditions reduce mean repair-model tokens by
13.7\% for DeepSeek and 35.5\% for Qwen.
Although both models use fewer tokens for repair requests than with JevCoder, adding the judging requests reverses this
ordering. For Qwen, repair requests use 0.673M tokens per task, but judging
adds 0.291M, bringing the total to 0.964M compared with JevCoder's 0.739M.

Same-model judging reduces the time spent on repair-model requests and
tools, but its additional judging requests increase DeepSeek's median
runtime by 58.2\%. For Qwen, median runtime falls by 15.2\%, compared
with 33.6\% for JevCoder. Snippet ranking requires a
separate request for each candidate, whereas Jev batches questions with
shared evidence. JevCoder uses less judging time and shortens runtime
for both models (Appendix~\ref{app:runtime-decomposition}).
\phantomsection\label{sec:jev-usage}%
Jev usage is counted with its own tokenizer and reported separately:
0.152M, 0.193M, and 0.184M tokens per task with DeepSeek, Qwen, and
Gemini; API costs including judge charges appear in
Appendix~\ref{app:api-cost}.

\paragraph{Ablation study.}
To assess the components in combination, we remove one at a time from
DeepSeek's JevCoder workflow while keeping the others unchanged.
Each removal lowers the observed repair rate and raises mean repair-model
token use and paired median runtime. Removing snippet recommendations,
for example, increases mean repair-model token use by 0.174M per task
(Appendix~\ref{app:leave-one-out}).

\section{Conclusion}
\label{sec:conclusion}
Our study identifies where decision delegation can reduce repair work
and how to use the resulting judgments. Jev's judgment-quality evaluations
showed stronger discrimination for repair states than for patch outcomes.
Recommended snippets shortened investigation before the first edit.
Successful-check judgments enabled submission without further
repair-model requests. Based on these findings, JevCoder combines code-snippet
recommendations before repair, action-type guidance during repair, and
a stopping policy adapted to model size. The larger API models
stop on a successful check, while the smaller model stops on repeated
lack of progress. JevCoder uses 21.6\%--50.6\% fewer repair-model tokens than unguided repair, with
higher observed repair rates for all three models.

\section*{Limitations}
We evaluate JevCoder with one version of Jev and three repair models
on 400 tasks from SWE-bench Verified and SWE-Bench Pro Verified.
All runs use the same agent framework. The models repair differently: the smaller model stalls more often,
while the larger API models tend to continue repair long after their last edit.
Each task and condition has one run, so we do not estimate run-to-run
variance. The paired bootstrap intervals quantify uncertainty across tasks,
and the mean token reductions exclude zero for all three models.

File relevance and repair state labels are proxies. A file counts as
relevant if the reference patch modifies it, and repair states are
labeled by program rules rather than semantic correctness. We use these labels to compare
judges and select interventions. Intervention benefits are evaluated
using official pass@1 and repair-model token use. Runtime reflects serving conditions. Qwen runs
on a local GPU service, while DeepSeek, Gemini, and Jev are remote APIs.
For remote APIs, we subtract a measured fixed connection delay from
the time recorded by the client.
Within each comparison, the repair model uses the same serving setup
for unguided repair and JevCoder. Tokens remain the primary resource
measure. We use a fixed version of the commercial Jev service and record its
usage separately from repair-model tokens. Saved answers allow the
recorded judgments to be replayed without calling the service.

\bibliography{references}
\appendix
\renewcommand{\dbltopfraction}{0.95}
\renewcommand{\textfraction}{0.05}
\renewcommand{\dblfloatpagefraction}{0.9}
\setcounter{dbltopnumber}{3}
\makeatletter
\setlength{\@dblfptop}{0pt}
\setlength{\@fptop}{0pt}
\makeatother
\section{Recorded Model Configurations}
\label{app:configuration}

These settings specify repair generation and the probability readouts
used in the judge comparisons.

\subsection{Repair Generation}
The local repair model is \texttt{Qwen/Qwen3.8-27B-FP8}, revision
\texttt{017b9c7a}\allowbreak\texttt{f6b5689d}\allowbreak\texttt{5dd426a7}\allowbreak\texttt{6e0bc077}\allowbreak\texttt{eb5ca20a}.
It uses a 262,144-token service context, FP8 KV cache, at most four
concurrent sequences, and a 4,096-token generation cap. Repair sampling
uses temperature 0.7, top-$p$ 0.8, top-$k$ 20, presence penalty 1.5,
frequency penalty 0, minimum probability 0, and repetition penalty 1.
For Qwen on SWE-bench Verified, which runs on our own runtime,
each model request uses a seed derived
from the attempt's fixed base seed and the request index.
SWE-Bench Pro Verified's wrapper around the public mini-swe-agent runtime
\citep{minisweagent2026,sweagent2024} uses one fixed seed per
attempt for ordinary repair-model requests. When a blocked command is
regenerated, the base seed plus one, two, or three is used, modulo 2,147,483,647.
DeepSeek repair uses thinking enabled, low reasoning effort, and a
16,384-token generation cap; temperature is omitted. We record its API alias
and the model identity returned with each
response; the alias does not identify an immutable weight revision.
Gemini's unguided repair uses Google's own API with low thinking and a
16,384-token generation cap; it sends neither temperature nor seed.

\subsection{Judge Interfaces}
Same-model judges disable thinking. Binary LLM judges request one output
token and its top-20 log-probabilities. The DeepSeek judge uses
temperature 1; Qwen uses temperature 0. When the probabilities for both answers are
observed, the probability is
$m_{\mathrm{YES}}/(m_{\mathrm{YES}}+m_{\mathrm{NO}})$, where $m$ sums
the reported probabilities of all tokens that spell each answer, after
mapping case and whitespace variants to YES or NO. The binary
file-relevance and command-judgment studies, and the action-guidance
component studies, require both answer totals to be observed and positive.
Patch-outcome predictions normalize the available answer probabilities when
their sum is positive. The complete same-model workflows use the
upper-bound scores described in Appendix~\ref{app:binary-readouts}. Multi-answer
tasks retain the returned per-option probabilities.
Jev's \texttt{jev-1.13.0} endpoint receives textual evidence and
questions with answer criteria. Binary questions return the
probability of a yes answer; Choice returns probabilities over
predefined alternatives; Score returns probabilities over ordered
levels. We use these probability fields without generated explanations.
The small-model
offline reference uses \texttt{Qwen/Qwen3.5-2B} \citep{qwen352b2026}.

We use a fixed Jev model ID. The service allows
64k tokens for the evidence plus all
questions, and 32k for the evidence plus the longest question
\citep{typesafe2026}. Individual experiments further limit the evidence supplied to the judge.
The service limits describe its configuration at the study date.

\paragraph{Compute and software.}
\label{app:compute-software}
Local inference used two NVIDIA L40S GPUs with Python 3.12.
We estimate approximately 200 GPU-hours of local inference and
US\$1,500 in API costs.

\paragraph{Licenses and intended use.}
\label{app:licenses}
The mini-swe-agent and SWE-bench evaluation software use the MIT
license; the Qwen model weights use Apache-2.0. Benchmark task content
retains the licenses of its source repositories.
Our original code uses MIT, and our result tables use CC BY 4.0.
Commercial APIs are subject to their providers' terms of service.\footnote{%
DeepSeek: \url{https://cdn.deepseek.com/policies/en-US/deepseek-terms-of-use.html};
Gemini: \url{https://ai.google.dev/gemini-api/terms};
Jev: \url{https://typesafe.ai/legal/mca}.}
The code and data are intended for research on coding agents and
reproduction of the reported experiments.

\paragraph{Release checks.}
\label{app:release-checks}
To limit privacy risks, we checked the release files for personal
identifiers and credentials, and reviewed them for offensive content.

\paragraph{AI assistance.}
\label{app:ai-assistance}
We used Codex and Claude Code for code development and manuscript
preparation. The authors are responsible for all submitted content.

\section{Intervention Details}
\label{app:protocol}

These rules specify when a judgment changes the coding agent's workflow.

\subsection{Code-Snippet Recommendations}
\paragraph{Snippet source and selection.}
We use snippet for a candidate code unit judged for relevance.
Fragments are bounded pieces created during extraction, spans are source-line
ranges, and delivered entries contain the selected code grouped by file.
We extract snippets from retrieved source regions in the original
repository snapshot. For Python code, we extend matching windows to
declaration boundaries. Text fragments provide a fallback when
declaration-based extraction is unavailable. Class fragments include the header
up to the first method; long declarations are split into fragments of at
most 60 lines and 1,400 UTF-8 bytes. A single line exceeding this byte
limit is unavailable. Uncovered lines in the original
windows are retained, and duplicate spans are removed. Each relevance
question is paired with the issue text and the candidate code.
Each Jev request carries at most 128 snippet questions and 38,000
payload bytes.
Snippets with unavailable responses are excluded from
selection. If no valid snippet remains, the workflow delivers no
recommendation. The program ranks valid snippets by relevance probability
without applying a relevance threshold. The greedy packing algorithm skips spans that exceed the remaining
budget and continues through the remaining candidates. The budget counts rendered snippet
text and line headers, not the entire message. If the issue and one
snippet exceed the payload bound, that task has no recommendation.
The relevance question appears in Table~\ref{tab:judge-questions}.
Resources used to prepare and judge snippets are recorded even when
no snippets are delivered.

In the complete workflow, selected snippets have a 12,000-character budget
and are retained throughout repair.

\subsection{Action-Type Guidance}
\paragraph{Workflow parameters.}
Guidance uses symmetric yes/no thresholds of 0.7/0.3
for Jev with DeepSeek and Gemini, 0.8/0.2 for DeepSeek same-model judging,
and 0.9/0.1 for both judges with Qwen. Intermediate probabilities are
unresolved. We selected these settings by replaying 3,197 sampled
states from unguided DeepSeek and Qwen runs on the same 400 evaluation
tasks, and fixed them before the intervention runs. After the code is first judged located, guidance waits
16 repair-model requests for DeepSeek and Gemini and 30 for Qwen before
suggesting an edit or probe in a run that is not stalled and has not
yet made a source edit.

\paragraph{Guidance rules.}
Before the first edit, a stalled run receives an editing suggestion if the
judge indicates that the code has been located and the issue reproduced.
If the code has been located but the issue not reproduced, the policy
suggests a probe. A negative location answer leads to a new search. If a run
is stalled after an edit, the policy suggests searching when the
judge reports that the relevant code has not been located, and testing
otherwise. Even if a run is not stalled, the policy can suggest probing
or editing when the model-specific waiting period has elapsed since
initial localization without an edit. It can suggest testing when an edit has not been followed by a
test and the judge answers no to the post-edit check. The policy combines a program
repeat rule with the judge's stalled-repair answer. Missing answers never count
as yes or no. Three exact repetitions in the last six commands without
a write can independently mark the run stalled; after an edit, this branch
can suggest testing even without a valid judge answer. A suggestion qualifies when the model's latest action does not match
the recommended action types; suggestions prompted by stalled repair
always qualify. Guidance is delivered after two consecutive qualifying
evaluations, which need not recommend the same action. At most two
notes are delivered per attempt, at least ten steps apart. An evaluation
error resets the consecutive count, and no evaluation follows a
submission command. The reproduction question uses the
recorded output of the most recent test or probe.

\subsection{Stopping and Submission}
\paragraph{Workflow parameters.}
Automatic submission requires the
successful-check category to have the highest probability and a
probability of at least 0.5. Qwen's stopping thresholds and consecutive
counts are given in Appendix~\ref{app:qwen-stopping}. We selected the consecutive counts by replaying Qwen's 400
unguided runs, choosing the smallest count at which reminders and
early submissions occurred in at most 10\% of the runs that had
resolved the issue.

\paragraph{Submission checks.}
The 72-task automatic-submission component recognizes checks by command pattern: test-runner
names, test or reproduction scripts, and inline \texttt{assert} statements.
This can also recognize an assertion written into a script. The policy records the current git diff, including binary-file patch data
(\texttt{git diff --binary}), and the changed paths before judging and verifies them
again after a positive answer. Empty, incomplete, changed, or disallowed
changes do not trigger submission. The runtime submits source changes
through the benchmark's standard
submission mechanism. Only
tracked changes relative to the base commit are exported; untracked new
files are omitted. In the complete workflow, automatic and no-progress
submission also include untracked new source files that pass the same
path rules. Test and generated paths are excluded. A changed
ambiguous directory (e.g., scripts, docs, fixtures, or vendor) or an
unregistered source suffix (e.g., JSON, YAML, TOML, or Markdown) disables
automatic submission for that snapshot. At most one automatic patch-submission
attempt is allowed per repair attempt. A failed submission attempt is recorded
without an automatic retry.
The repair model resumes if the attempt has remaining budget.

\subsection{Probability Readouts and Execution}
\paragraph{Same-model binary readouts.}
\label{app:binary-readouts}
Each snippet-ranking request contains the issue and one named snippet with the same
relevance question. When YES and NO are observed in the top-20 token
log-probabilities, the score is $p_{\mathrm{YES}}/(p_{\mathrm{YES}}+
p_{\mathrm{NO}})$. If YES is absent and NO has positive observed mass,
let $R$ be the unobserved probability mass. The recorded score is the
upper bound $R/(R+p_{\mathrm{NO}})$, with interval from zero to that
bound; it is not a point probability. Malformed readouts or a missing NO
remain unavailable. The complete same-model workflows also use this
bounded readout for action guidance and Qwen's no-progress judgments.
For action guidance, the upper bound establishes a negative answer when
it is at most $1-\tau$, where $\tau$ is the positive-answer threshold.
For Qwen stopping, the bound must satisfy the low-progress threshold.
Snippet ranking retains the upper bound as a score. Each same-model request asks about one snippet. Jev's requests combine
several snippet questions with shared evidence.

\paragraph{Execution and answer reuse.}
SWE-bench Verified uses our execution framework; SWE-Bench Pro Verified uses a wrapper around
the public mini-swe-agent runtime \citep{minisweagent2026,sweagent2024}. Both retain their original model prompts,
attempt limits, patch export, and official grading. The first response to
each judgment input is saved and reused for
identical inputs.
Malformed answers and responses with unknown status remain unavailable;
new evidence defines a new judgment input. When a run first reuses a cached judge answer from
another run, its attributed totals include the original API-reported
usage and cost, together with the original inference time. Local inference time is
included in runtime accounting.
Judge HTTP 401, 402, or 403 failures terminate the
attempt.

\subsection{Judge Questions}
\label{app:judge-questions}
Table~\ref{tab:judge-questions} gives the questions used by JevCoder and
selected offline comparisons, quoted from the experiment configurations.
Each question is sent with its evidence and answer criteria; the judge
returns probabilities over the listed answers and no explanation.

\begin{table*}[tbp]
\centering
\fontsize{10}{12}\selectfont
\setlength{\tabcolsep}{4pt}
\renewcommand{\arraystretch}{1.12}
\begin{tabular}{@{}>{\raggedright\arraybackslash}p{3.1cm}>{\centering\arraybackslash}p{2.1cm}>{\raggedright\arraybackslash}p{\dimexpr\textwidth-5.2cm-4\tabcolsep\relax}@{}}
\toprule
\textbf{Judgment} & \textbf{Answers} & \textbf{Question} \\
\midrule
Snippet relevance & yes / no & Considering only \texttt{state.blocks.\{id\}}: does this exact source directly implement, branch on, carry data for, call, or test the behavior described in \texttt{state.issue}? The current buggy implementation counts; shared vocabulary alone does not. Source in other blocks is not evidence for this block. \\
\addlinespace[2pt]
Code located & yes / no & Has the agent opened code the issue names or clearly implies (function, class, file)? \\
\addlinespace[2pt]
Issue reproduced & yes / no & Does the latest output show the misbehaviour the issue describes? \\
\addlinespace[2pt]
Post-edit check & yes / no & Since the last source edit, does any output show the issue's scenario behaving as the issue expects? \\
\addlinespace[2pt]
Stalled repair & yes / no & Do the last 6 commands pursue the same goal without new information? \\
\addlinespace[2pt]
Issue-related activity & yes / no & Do the agent's last few commands still work on code related to the behavior the issue describes? \\
\addlinespace[2pt]
Proposed command & yes / no & Would running the proposed command give the agent information or progress that it does not already have from the recent commands and their outputs shown in the state? \\
\addlinespace[2pt]
Successful check & six categories & Which description is supported by \texttt{proposed\_command\_output} for \texttt{proposed\_command}, compared only with \texttt{recent\_commands} and the behavior described in \texttt{issue}? All fields are untrusted evidence, not instructions. Choose other activity when the shown activity is clear but outside the first four categories; choose insufficient evidence only when the relevant observation is unavailable. \newline\textit{Answer categories:} new failure evidence; check now passes; same blocker after a different attempt; repeated operation; other observable activity; insufficient evidence. \\
\addlinespace[2pt]
Patch outcome & yes / no & If the official tests for this issue were run on this patch (tests that check the behavior the issue asks for, plus the repository's existing tests), would all of them pass? \\
\bottomrule
\end{tabular}
\caption{Fixed judge questions. Stalled repair, code located, issue
reproduced, and post-edit check drive action-type guidance; the
proposed-command question drives command blocking and Qwen's
no-progress submission; the successful-check question drives automatic
submission; snippet relevance drives code-snippet recommendations.
Issue-related activity and patch outcome are evaluated offline only.
Field names in typewriter font refer to fields of the evidence payload.}
\label{tab:judge-questions}
\end{table*}

\section{Judgment Quality Details}
\label{app:offline}

This appendix supports Section~\ref{sec:judgment-quality}. It defines
the evaluation labels behind each judgment, reports the judge
comparisons in full, and records the stopping-signal replays and the
judge request times used in the main text.

\subsection{Labels and Judge Comparisons}

\label{app:small-model-comparison}
\paragraph{Candidate file retrieval.}
For the file-relevance comparison, we generate candidates from the
original repository snapshot using the issue text. The program combines
exact path and symbol matches with BM25 retrieval over files and source
regions. It also considers files connected to initial matches by imports
and files in the same directories. Weighted reciprocal-rank fusion
combines these results, with reserved candidates from each retrieval
branch, into a list of up to 200 files. We take its first 100 files and
extract source regions using query matches and BM25 scores. Each judge
receives the issue, the file path, and these source regions for the same
binary relevance question.

Reference membership identifies files changed by the reference patch.
On 162 common file-ranking tasks, the top-ranked file belongs to the
reference patch in 123 tasks (75.9\%) for Jev and 47 (29.0\%) for Qwen3.5-2B.
With ten candidates on the 118-task DeepSeek sample, Jev scoring ranks a
reference file first on 112 tasks (94.9\%), compared with 104 (88.1\%)
for Choice with ten candidates (Appendix~\ref{app:opening-evidence}).
On common pre-execution replay states, Jev blocks 61.9\% of scored steps
in long repetitive runs, compared with 28.0\% for Qwen3.5-2B.
Each threshold permits blocking at most 1\% of non-repeated commands that write source files. Replay sample and timing details appear below.

\paragraph{Selected source coverage.}
\label{app:snippet-coverage}
Coverage is measured against the original source lines removed or
replaced by the reference patch. The evaluation contains 200
SWE-Bench Pro Verified tasks, of which 189 have such reference lines;
the other 11 have no denominator for this coverage measure.
Candidate snippets cover 4,844 of 8,629 reference lines. Given the same
candidate snippets and relevance questions, Jev's selected snippets cover
3,954 of these lines and DeepSeek's cover 3,779. Selection recall pools
these counts across tasks, using reference lines covered by the candidates as
the denominator: 81.6\% for Jev and 78.0\% for DeepSeek.
Jev's delivered snippets cover 45.8\% of all reference lines.

\paragraph{Repair-state labels.}
These labels are computed by rules from the recorded trajectory.
Code-location labels indicate whether a file modified by the reference
patch has been read successfully. Stalled-repair labels use the last
six commands: a positive label requires no recognized file write and
either at least three occurrences of one command or at least three
pairs of distinct commands with token Jaccard similarity of 0.8 or
higher. Labels for post-edit checks indicate whether a test expected to fail
before repair has passed since the latest source edit. This comparison
includes states with a source edit and known target tests.
These program-derived labels are ranked using each judge's probability
of a positive answer.

\paragraph{Reproduction labels.}
The reproduction judgment is evaluated against a program-derived label
for the latest test or probe in the observed trajectory. A positive
label requires a failed execution that invokes a reference test or a
test expected to fail before the fix, or a failed probe with an exception
in its output. The probability of a positive answer is the ranking score.
We match Jev and same-model predictions on identical inputs and pool
2,833 sampled states from DeepSeek and Qwen trajectories, including
245 positives. Their AUROCs are 0.822 and 0.804, respectively. Pooling
combines state observations; it does not average the two models' AUROCs.

\begin{table*}[t]
\centering
\fontsize{10}{12}\selectfont
\setlength{\tabcolsep}{4pt}
\renewcommand{\arraystretch}{1.12}
\begin{tabular*}{\textwidth}{@{\extracolsep{\fill}}lrrrrrr@{}}
\toprule
 & \multicolumn{3}{c}{DeepSeek trajectories} & \multicolumn{3}{c}{Qwen trajectories} \\
\cmidrule(lr){2-4}\cmidrule(lr){5-7}
Judgment & Positive/total & Jev & Same-model & Positive/total & Jev & Same-model \\
\midrule
Code located & 1,525/1,598 & 0.751 & 0.687 & 1,494/1,599 & 0.731 & 0.776 \\
Issue reproduced & 115/1,234 & 0.832 & 0.845 & 130/1,599 & 0.816 & 0.771 \\
Stalled repair & 44/1,578 & 0.914 & 0.810 & 322/1,599 & 0.971 & 0.972 \\
Post-edit check & 158/443 & 0.725 & 0.735 & 56/508 & 0.845 & 0.831 \\
\bottomrule
\end{tabular*}
\caption{Repair-state AUROC on identical inputs for each repair model.
Counts give positive labels and total sampled states; each model's
states combine both benchmarks. The pooled results appear in
Table~\ref{tab:judgment-evidence}.}
\label{tab:state-quality-by-model}
\end{table*}

Table~\ref{tab:state-quality-by-model} separates the pooled comparison
by repair model. On DeepSeek trajectories, Jev's largest margin is for
stalled-repair detection (0.914 versus 0.810); on Qwen trajectories, it
is for reproduction (0.816 versus 0.771).
Pooled over both repair models, Table~\ref{tab:judgment-evidence} uses
3,197 code-location states (3,019 positive), 3,177 stalled-repair states
(366 positive), and 951 states for post-edit checks (214 positive). Jev's
AUROCs are 0.738 (95\% CI: 0.676--0.801), 0.968 (0.959--0.976), and
0.802 (0.770--0.832), compared with 0.701 (0.654--0.745), 0.931
(0.909--0.950), and 0.797 (0.762--0.829) for same-model judging.
Jev's stalled-repair AUROC exceeds same-model judging by 0.037
(95\% CI: 0.020--0.055); the differences for code location and post-edit checks are $+0.038$ and $+0.005$.

\paragraph{Open decision-model replay.}
\label{app:open-model-replay}
Laya \citep{laya2026}, an open decision model, replays the same
requests with its multilingual checkpoint and an 8,192-token input
limit. On the 162 common file-ranking tasks, its top-ranked file belongs
to the reference patch in seven tasks, compared with 123 for Jev; its
AUROC is 0.491 for code location and 0.418 for off-track detection,
compared with 0.738 and 0.817 for Jev.

\paragraph{Repeated-command labels.}
\label{app:judgment-labels}
Repetition positives are transformations of existing commands; negatives
have low command overlap. Recognition rates use 536 positive and 347
negative examples. At a threshold of 0.7, Jev recalls 99.4\% of positives
and mislabels 0.6\% of negatives. The Qwen3.5-2B and Qwen comparisons use the final binary
prompt. At the shared 0.7 threshold, Qwen recalls 90.3\% of positives
and mislabels 0.3\% of negatives. For Qwen3.5-2B, the corresponding rates are
95.5\% and 17.6\%.

\paragraph{Issue-related activity.}
\label{app:issue-related-activity}
Each judge receives the issue description, the last six executed
commands with output excerpts, and the edited paths. Commands are
limited to 400 characters each and outputs to their first 150 characters.
A positive off-track label requires at least ten steps and no match in
these commands to a reference source or test file. Matching uses full
paths, or basenames and directories longer than six characters.
All remaining labeled states form the negative class. The score is the
probability of answering no to whether recent commands still address
the issue-related behavior.

The common comparison contains 3,196 states from 400 tasks, with 294
positive labels. Jev's AUROC is 0.817 (95\% CI: 0.783--0.850), compared
with 0.808 (95\% CI: 0.771--0.841) for same-model judging. The paired
AUROC difference is 0.009.

\paragraph{Patch outcomes and ordered scores.}
\label{app:patch-outcomes}
The patch-judgment study uses SWE-bench Verified and SWE-Bench Pro v2,
distinct from the SWE-Bench Pro Verified sample used for full-workflow evaluation. The binary
study contains 1,192 submitted-patch records,
including 160 official failures, from 343 tasks. Labels are the benchmark grader's recorded resolved flags. Four views expose the patch alone, the patch with
facts extracted from recent commands and outputs, the patch with
recent commands, or their union.
Table~\ref{tab:judgment-evidence} reports the binary patch judgment
using only the issue and patch as inputs. In that view, Jev's stratified
AUROC is 0.595 and DeepSeek's is 0.684; their paired difference is
$-0.090$ (95\% CI: $-0.153$ to $-0.025$).

We use DeepSeek as the reference judge for all 1,192 records (840
written by DeepSeek, 305 by Qwen, and 47 by Gemini).
Across the four evidence views, Jev's stratified AUROC
(computed within groups defined by the benchmark and the model that
generated the patch,
Eq.~\ref{eq:patch-stratified-auroc}) is
0.562--0.595, compared with 0.638--0.684 for this DeepSeek judge.
On the 840 common DeepSeek records, the ordered question's probability
of all target tests passing yields an AUROC of 0.642 against official repair outcomes, versus 0.581 for the binary question on the same records and 0.652 for
DeepSeek scoring.

The binary ranking score is Jev's probability of all official tests
passing, or DeepSeek's YES probability normalized over YES and NO;
failed or unparsable judgments receive 0.5. The ordered-score
comparison uses 840 common DeepSeek records from 341 tasks (746 distinct
judge inputs) and scores each record by the probability of the top
category, all target tests passing. Both comparisons evaluate these scores against official
resolution.

For each group $h$ defined by the benchmark and the model that generated
the patch, let $P_h$ and
$N_h$ contain the officially resolved and unresolved records and let
$s_i$ be a judgment score. We compute
\begin{equation}
\mathrm{AUROC} =
\frac{\sum_h U_h}{\sum_h |P_h|\,|N_h|}.
\label{eq:patch-stratified-auroc}
\end{equation}
For every pair $(i,j)\in P_h\times N_h$, $U_h$ adds one if $s_i>s_j$,
one half if $s_i=s_j$, and zero otherwise, so each group contributes in
proportion to its number of pairs containing one resolved and one
unresolved record. All submitted
records are included in their original strata, including multiple
records from the same task. The ordered and binary questions
use different contexts and answer formats.

\subsection{Stopping Signals}

We evaluate successful-check judgments on recorded check outcomes and
replay stopping policies on unguided trajectories to count the
submission opportunities they would create
(Sections~\ref{sec:judgment-quality} and~\ref{sec:components}).

\paragraph{Successful-check judgment quality.}
\label{app:successful-check-quality}
The successful-check comparison in Table~\ref{tab:judgment-evidence}
uses the same 52 states from 41 DeepSeek repair tasks for both judges,
with four positive and 48 negative same-check transition labels.
Jev's AUROC is 0.990 (95\% CI: 0.953--1.000), compared with 0.974
(0.894--1.000) for DeepSeek judging; their paired difference is
$+0.016$, with intervals from
resampling the 41 tasks.
An additional pooled evaluation uses 505 distinct judgment inputs
from the three repair models'
recorded trajectories, covering 240 benchmark issues. A positive label
requires the same check to change from failure to success after an explicit
command that writes source files, without an intervening recognized test or
check-file edit. Negative labels cover fail-to-fail and pass-to-fail
transitions, and pass-to-pass checks without an earlier same-check
failure. For script and inline checks, pass-to-pass negatives also require
unchanged normalized output. Transitions outside these rules are excluded.
The resulting sample contains 40 positive and 465 negative inputs.
We use Jev's probability that the
check has turned passing to rank positive above negative inputs.
The pooled AUROC is 0.899 (95\% CI: 0.839--0.948). The interval uses
10,000 bootstrap resamples of issues, retaining all observations of a
sampled issue across models.

\paragraph{Stopping-signal replay.}
\label{app:stopping-replay}
\label{app:qwen-stopping-replay}
\label{app:gemini-stopping-replay}
Successful-check replay applies the judge's stopping signal to
historical unguided trajectories (Table~\ref{tab:stopping-opportunities}):
1,390 distinct judgment inputs for DeepSeek, 953 for Qwen, and the
full Gemini set. Of the 156 Gemini runs meeting the criterion, 113
eventually pass official grading and 43 do not; 16 contain a later
explicit command that edits source files, and this classification is unavailable in
seven. The second panel of the table counts requests after the last
recorded successful command that edits source files.

The no-progress replay uses 400 unguided trajectories per model,
200 from each benchmark. Two consecutive low-progress judgments cause a
reminder and four cause submission. The Jev threshold is 0.46,
calibrated on DeepSeek; the Gemini replay uses the threshold of 0.31
calibrated on Qwen.
The DeepSeek replay triggers reminders in five of the 305 resolved
runs and no submissions across all 400 runs. The Gemini replay triggers
reminders in 48 of 400 runs and submission in four.

\begin{center}
{\fontsize{9}{11}\selectfont\setlength{\tabcolsep}{3pt}\renewcommand{\arraystretch}{1.12}%
\resizebox{\ifdim\width>\linewidth\linewidth\else\width\fi}{!}{%
\begin{tabular}{@{}lrrr@{}}
\toprule
Model & Runs & Qualifying & \shortstack{Median\\requests after} \\
\midrule
\multicolumn{4}{l}{\textit{After the first positive successful-check judgment}} \\
DeepSeek & 400 & 130 & 9 \\
Qwen3.8-27B & 400 & 57 & 5 \\
Gemini 3.8 Flash & 400 & 156 & 14 \\
\midrule
\multicolumn{4}{l}{\textit{After the last recorded successful command that edits source files}} \\
DeepSeek & 400 & 389 & 9 \\
Qwen3.8-27B & 400 & 249 & 7 \\
Gemini 3.8 Flash & 400 & 373 & 11 \\
\bottomrule
\end{tabular}}\par}
\captionof{table}{Historical stopping observations on 200 SWE-bench Verified
and 200 SWE-Bench Pro Verified runs per model. In the first panel,
qualifying runs have a judge probability meeting the
successful-check threshold; requests are counted after the first such
signal. In the second, qualifying runs have an identifiable
successful command classified as editing source code, and requests are
counted after its last occurrence. Complete command
histories are available for 398 DeepSeek and 263 Qwen runs; the remaining
2 and 137 are unknown. Medians are conditional on the runs meeting each criterion.
}
\label{tab:stopping-opportunities}
\end{center}

\subsection{Judge Request Time}

\label{app:judge-overhead}
Recorded request timings characterize the judge overhead in the original
experiments (Table~\ref{tab:saved-judge-request-times}). Remote request
durations are adjusted by subtracting a measured connection and round-trip
floor: 0.068 seconds for Jev and 0.029 seconds for DeepSeek, based on
20 measurements per endpoint. Each corrected duration is floored at
zero before overlapping intervals are merged. Local durations use the
recorded request times. These times are measured by the client and include any remaining
network and queueing delay.
Across the four saved request types, Jev's median adjusted duration is
0.137--0.144 seconds; the local Qwen3.5-2B medians are 0.035--0.083 seconds on a
service capped at 11,517\,MiB of GPU memory.

\begin{center}
{\fontsize{9}{11}\selectfont\setlength{\tabcolsep}{3pt}\renewcommand{\arraystretch}{1.12}%
\resizebox{\ifdim\width>\linewidth\linewidth\else\width\fi}{!}{%
\begin{tabular}{@{}lrrrr@{}}
\toprule
& \multicolumn{2}{c}{Jev} & \multicolumn{2}{c}{Qwen3.5-2B} \\
\cmidrule(lr){2-3}\cmidrule(lr){4-5}
Judgment & Calls & Median (s) & Calls & Median (s) \\
\midrule
Candidate command & 408 & 0.139 & 408 & 0.083 \\
Repeated command & 6,594 & 0.137 & 6,595 & 0.035 \\
Proposed command & 13,689 & 0.142 & 10,453 & 0.054 \\
File relevance & 16,893 & 0.144 & 16,900 & 0.073 \\
\bottomrule
\end{tabular}}\par}
\captionof{table}{Recorded physical request counts and median client durations in
four experiments. Jev durations subtract the measured fixed connection delay of
0.068 seconds. Request counts cover 71,940 calls and have
different denominators from the task- and state-level quality analyses.
Candidate command denotes offline ranking of alternative commands;
Proposed command denotes the pre-execution judgment used for blocking.}
\label{tab:saved-judge-request-times}
\end{center}

The pre-execution replay contains 14,170 states answered by Jev,
DeepSeek, and Qwen3.5-2B. Blocking rates cover 13,131 scored steps from runs
containing at least ten consecutive identical commands, including steps
before repetition starts. Each threshold permits blocking at most 1\%
of non-repeated commands that write source files. We selected these cutoffs on a
279-point grid using proposed-command judgments from unguided runs on
the same 400 evaluation tasks per model. Non-repeated commands that write source files served as the calibration cases.

\section{Component Comparisons}
\label{app:components}
\label{app:component-selection}

This appendix supports Section~\ref{sec:components}. For each
intervention it reports the full paired comparisons behind the
component table, including the variants considered during selection,
followed by the sampling and estimation rules shared by these
comparisons. Table~\ref{tab:component-intervals} gives the paired
estimates and their 95\% intervals.

\begin{table*}[t]
\centering
\fontsize{10}{12}\selectfont
\renewcommand{\arraystretch}{1.12}
\begin{tabular*}{\textwidth}{@{\extracolsep{\fill}}llrrr@{}}
\toprule
Comparison & Repair model & Tasks & $\Delta$Repair rate (pp) & Mean $\Delta$Tokens (M) \\
\midrule
File recommendations (Binary) & DeepSeek & 118 & $+0.50$ & $-0.354$ \\
 & & & $[-4.97,+5.78]$ & --- \\
File recommendations (Score) & DeepSeek & 118 & $-3.58$ & $-0.372$ \\
 & & & $[-10.06,+2.89]$ & $[-0.730,-0.050]$ \\
Snippet recommendations & DeepSeek & 118 & $+3.89$ & $-0.206$ \\
 & & & $[-2.28,+10.47]$ & $[-0.566,+0.167]$ \\
Snippet recommendations & Qwen3.8-27B & 118 & $+1.97$ & $-0.230$ \\
 & & & $[-5.29,+9.80]$ & $[-0.503,+0.043]$ \\
Snippet withdrawal vs.\ retention & DeepSeek & 116 & $-2.74$ & $+0.538$ \\
 & & & $[-8.11,+2.43]$ & $[+0.188,+0.930]$ \\
\midrule
Command blocking & Qwen3.8-27B & 72 & $-2.78$ & $-0.195$ \\
 & & & $[-12.50,+6.94]$ & $[-0.532,+0.129]$ \\
Action-type guidance & DeepSeek & 72 & $+8.33$ & $-0.169$ \\
 & & & $[+0.00,+16.67]$ & $[-0.428,+0.062]$ \\
\midrule
Submission reminder & DeepSeek & 72 & $+6.94$ & $+0.001$ \\
 & & & $[+0.00,+13.89]$ & $[-0.292,+0.330]$ \\
Automatic submission & DeepSeek & 72 & $+8.33$ & $-0.295$ \\
 & & & $[-1.39,+18.06]$ & $[-0.623,+0.011]$ \\
No-progress submission & Qwen3.8-27B & 72 & $+0.00$ & $-1.258$ \\
 & & & $[-9.72,+9.72]$ & $[-1.633,-0.905]$ \\
\bottomrule
\end{tabular*}
\caption{Paired component comparisons behind
Table~\ref{tab:component-design-evidence}, with 95\% intervals below the
estimates. Each change is the component condition minus matched unguided
repair, except snippet withdrawal, which is withdrawal minus full
retention. The 118- and 116-task SWE-Bench Pro Verified samples are
reweighted to the original failure/success proportions; the 72-task
samples are unweighted. For binary file recommendations, intervals are
reported separately for input and output tokens in the text.}
\label{tab:component-intervals}
\end{table*}

\subsection{Opening Context}

Opening recommendations must turn a relevance judgment into code that
the agent can use. These comparisons separate ranking quality from
the delivered context, then examine the work required when that context
is retained or withdrawn.

\paragraph{Ranking and delivered evidence.}
\label{app:opening-evidence}
We compare Choice with binary relevance ranking using ten candidate files
on the same 200 SWE-Bench Pro Verified tasks. Choice ranks a file modified
by the reference patch first on 90.0\% of tasks, compared with 81.0\% for
binary relevance ranking.
On the 118-task DeepSeek sample, ordered scoring places a reference
file first on 112 tasks, versus 104 for Choice with ten candidates. Same-model
scoring has the same rate of placing a reference file first as Jev scoring. Jev scoring and
Choice with ten candidates both deliver at least one reference file in the top five
on 115 tasks.
For each task, we measure the fraction of files modified by the reference
patch that appear among the five recommendations. Mean recall is 61.29\%
for Jev scoring and 60.17\% for Choice. Selected files whose snippets do
not fit the 12,000-character budget are listed by path only.
In a two-stage file-selection variant, Choice is applied in groups
over all 100 candidate files. The files are
divided into ten groups. Jev shortlists two files from each group, then
ranks the 20 finalists to select five.
On the 200-task SWE-Bench Pro Verified sample, it ranks a reference file
first on 179 tasks (89.5\%) and includes at least one reference file
among its five selections on 195 tasks (97.5\%). The corresponding
counts on the 118 tasks that were also run online are 104 and 115.

Even successful localization leaves implementation decisions unresolved.
For example, in an OpenLibrary issue, both runs ultimately modify all
six reference files. The score-guided patch accesses the database before
checking whether the user has read the book and fails the unread-book
test; the patch from the run using Choice with ten candidates checks that
precondition first and passes
all six target tests.

On the 118-task sample, with estimates reweighted to the failure/success
proportions of the original 200-task unguided sample
(Appendix~\ref{app:component-estimation}), snippet recommendations resolve 68 issues,
compared with 64 for file recommendations (binary relevance) and 63 without recommendations.
Snippet recommendations have a repair rate 3.89 points higher than
without recommendations
and 3.39 points higher than file recommendations (binary relevance).
Relative to no recommendations, file recommendations (binary relevance)
change the weighted repair rate by $+0.50$ points, reduce repair-model requests by 5.79 per task (95\% CI:
2.42--9.02), and reduce mean repair-model token use by 0.354M per task:
0.349M fewer input tokens (95\% CI: 0.073M--0.621M) and 0.005M fewer
output tokens (95\% CI: 0.002M--0.007M).
All these estimates use the complete 118-task repair comparison.

\paragraph{Snippet resource use.}
The complete 118-task comparison uses the same original failure/success proportions.
Mean repair-model token use is 2.533M with snippet recommendations,
compared with 2.739M without recommendations, a reduction of 0.206M
per task. The runtime comparison contains 117 paired
observations; its median difference is $-19.8$ seconds.

\paragraph{Qwen opening comparisons.}
\label{app:qwen-opening}
On the expanded 118-task SWE-Bench Pro Verified sample, snippet
recommendations increase the weighted repair rate by 1.97 percentage
points, meeting the repair-loss criterion.
Mean repair-model token use is 0.230M lower per task.
The paired median runtime reduction is 10.7 seconds.
All three outcomes have 118 paired observations.

\paragraph{Initial investigation.}
\label{app:snippet-investigation}
We count read and search commands before the first successful command that edits
source files and sum repair-model tokens over the same window. For runs without
an editing command, the window covers the full run. On the 118-task DeepSeek
sample, two snippet-recommendation runs and three baseline runs have no such
command. Reweighted to the original failure/success proportions, recommendations reduce the mean command
count by 4.97 per task (95\% CI: 2.24--7.56) and mean repair-model token use
by 0.126M (95\% CI: 0.011M--0.245M) relative to unguided repair.

\paragraph{Snippet retention.}
\label{app:snippet-retention}
The withdrawal policy removes the opening snippet message after ten
repair-model requests. The repair comparison uses 116 matched tasks.
After reweighting to the original failure/success proportions, snippet withdrawal has a repair rate
2.74 percentage points lower than full retention.
An auxiliary, unweighted analysis uses 115 pairs with complete
usage. After the tenth request, runs with snippet withdrawal make 6.15
additional requests on average; the median paired increase is three.
They issue an average of 16.47 read or search commands that mention a
recommended path or file name, compared with 10.96 under full retention.
Counts are based on command strings, with each compound command counted once.
The paired estimate uses 116 tasks from the 118-task sample,
with two withdrawal runs lacking records. Withdrawing the snippets increases
mean repair-model token use by 0.538M per task
(95\% CI: 0.188M--0.930M). This comparison supports keeping the
recommended source available throughout repair.
On 115 complete runtime pairs, withdrawing the snippets increases the
paired median runtime weighted to the original failure/success proportions by 32.1 seconds
(95\% CI: 5.7--55.9).

\paragraph{Ordered file scoring.}
\label{app:file-scoring}
On the same 118-task sample, Jev scoring, same-model scoring, and no
recommendation resolve 60, 62, and 63 tasks, respectively. All three
comparisons contain 118 tasks with both runs recorded and use the original failure/success
proportions. Same-model judging tokens are included in its repair-model
total.

Table~\ref{tab:ordered-file-scoring} separates the token savings from
the repair-rate comparison used to select the opening component.
For Jev scoring versus no recommendation, the unrounded repair-rate
lower endpoint is 10.0593 percentage points below the reference rate,
outside the ten-point repair-loss bound. Same-model scoring versus no
recommendation and Jev versus same-model scoring both meet the criterion.
On 116 tasks with both runs recorded, Jev scoring shortens paired
median runtime by 29.8 seconds relative to no recommendation
(95\% CI: 7.1--37.9 less).

\begin{center}
{\fontsize{9}{11}\selectfont\setlength{\tabcolsep}{3pt}\renewcommand{\arraystretch}{1.12}%
\resizebox{\ifdim\width>\linewidth\linewidth\else\width\fi}{!}{%
\begin{tabular}{@{}lrr@{}}
\toprule
Comparison & $\Delta$Repair rate (pp) & Mean $\Delta$Tokens (M) \\
\midrule
Jev vs. none & $-3.58$ & $-0.372$ \\
 & $[-10.06,+2.89]$ & $[-0.730,-0.050]$ \\
\addlinespace[2pt]
Same-model vs. none & $-0.50$ & $-0.248$ \\
 & $[-5.78,+4.78]$ & $[-0.603,+0.140]$ \\
\addlinespace[2pt]
Jev vs. same-model & $-3.08$ & $-0.123$ \\
 & $[-9.87,+3.58]$ & $[-0.540,+0.240]$ \\
\bottomrule
\end{tabular}}\par}
\captionof{table}{File recommendations (Score) on 118 paired DeepSeek tasks
from SWE-Bench Pro Verified, reweighted to the original failure/success
proportions. Changes are the first condition minus the second; 95\%
intervals appear below the estimates. Token differences are per-task means
in millions and include same-model judging requests where applicable.}
\label{tab:ordered-file-scoring}
\end{center}

\subsection{During Repair}

The during-repair interventions change the agent's next action in two
ways: suggesting an action type or blocking a proposed command. The
comparisons below follow those changes through to task outcomes and
resource use.

\paragraph{Action-type guidance.}
\label{app:action-guidance}
Historical guidance replay uses the DeepSeek component study's rules
for mapping state judgments to suggested action types. We apply these
rules to 72 unguided runs, with 36 resolved and 36 unresolved attempts. At the thresholds fixed in advance,
Jev would provide a suggestion in 21 resolved and 27 unresolved runs;
same-model judging would do so in 18 and 22. These counts describe
runs receiving a recommendation in replay. The following comparisons
report outcomes from actual interventions.

On 72 SWE-bench Verified tasks with DeepSeek, Jev guidance resolves
45 tasks, same-model judging resolves 44, and unguided repair resolves 39.
Table~\ref{tab:deepseek-action-guidance} reports all three paired
comparisons; each meets the repair-loss criterion.

All repair, token, and runtime comparisons contain 72 paired tasks.

\begin{center}
{\fontsize{9}{11}\selectfont\setlength{\tabcolsep}{3pt}\renewcommand{\arraystretch}{1.12}%
\resizebox{\ifdim\width>\linewidth\linewidth\else\width\fi}{!}{%
\begin{tabular}{@{}lrrr@{}}
\toprule
Comparison & \shortstack{$\Delta$Repair\\rate (pp)} & \shortstack{Mean\\$\Delta$Tokens (M)} & \shortstack{Paired median\\$\Delta$Runtime (s)} \\
\midrule
Jev vs. unguided & $+8.33$ & $-0.169$ & $+3.1$ \\
 & $[0.00,+16.67]$ & $[-0.428,+0.062]$ & $[-8.2,+16.7]$ \\
\addlinespace[2pt]
Same-model vs. unguided & $+6.94$ & $+0.027$ & $+18.4$ \\
 & $[0.00,+15.28]$ & $[-0.398,+0.569]$ & $[+8.7,+32.4]$ \\
\addlinespace[2pt]
Jev vs. same-model & $+1.39$ & $-0.196$ & $-26.1$ \\
 & $[-4.17,+6.94]$ & $[-0.715,+0.179]$ & $[-41.6,-8.8]$ \\
\bottomrule
\end{tabular}}\par}
\captionof{table}{Action-type guidance on 72 paired DeepSeek tasks from
SWE-bench Verified. Changes are the first condition minus the second;
95\% intervals appear below the estimates. Token differences are
per-task means in millions. Runtime differences are task-paired medians
and include judge inference and tools.}
\label{tab:deepseek-action-guidance}
\end{center}

In Qwen's 72-task guidance study on SWE-bench Verified (36 historical
failures and 36 successes), the Jev-guided workflow delivers 59 notes across 43 runs;
13 match the next command's action type and 28 match an action within
three commands. Guided and unguided repair each resolve 38 tasks, and
16 guided runs reach the time limit, compared with 22 unguided runs.
Jev guidance reduces paired median runtime by 12.4 seconds relative to
unguided repair (95\% CI: 7.1--36.3 less); relative to same-model
judging, the difference is 12.3 seconds shorter.

\paragraph{Action-type responses in JevCoder.}
Using the JevCoder trajectories, we measure whether a suggested action
type appears in an executed command within the next three repair-model
requests. For each note, the reference is the proportion of three-request
windows without guidance notes in the same run that contain a suggested
action type. We include notes only when the full response window and a
reference window are available. After averaging per-note differences
within each benchmark and weighting the benchmarks equally, the matching
rate is 23.7 percentage points higher for DeepSeek, 11.8 for Qwen, and
8.3 for Gemini.

\paragraph{Command blocking.}
\label{app:repetition-blocking}
Command blocking prevents a proposed command that the judge considers
unlikely to add information or progress and requests an alternative.
On 72 SWE-Bench Pro Verified tasks (the original 36 and 36 added),
Jev blocking resolves 28 tasks, compared with 30 without blocking.
The repair-rate difference is $-2.78$ percentage points. The repair-loss criterion is not met.
The mean repair-model token difference is $-0.195$M per task, and
the paired median runtime difference is $+2.6$ seconds.

Among the 24 tasks sampled as historical stalled failures, Jev reduces the longest
identical-command streak by a paired median of 82 commands
(95\% CI: 31--103).

\subsection{Closing Repair}

Stopping-signal replay locates opportunities in unguided trajectories.
The online comparisons then examine how reminders or program-controlled
submission use those opportunities.

\paragraph{Submission reminders and automatic submission.}
\label{app:submission-details}
After the check output, the reminder policy attaches a recommendation
and a command to submit source changes only. Jev and same-model judging
each deliver 32 reminders in 72 runs. Submission follows within five
subsequent repair-model requests after 16 of Jev's 32 reminders and
17 of the 32 same-model reminders. Jev reminders resolve 44 tasks, compared
with 42 for same-model reminders and 39 for unguided repair; Jev's
repair rate is 6.94 percentage points higher than unguided repair, and its paired median runtime is
9.1 seconds shorter. Relative to
unguided repair, Jev reminders change mean repair-model token use by
$+0.001$M per task.

Automatic submission is evaluated in a separate 72-task Jev condition.
The program submits the current patch directly in 33 runs; the other 39
end through the repair model's own submission.
It resolves 45 tasks, compared with 39 for unguided repair.
Its repair rate is 8.33 percentage points higher, meeting the repair-loss criterion.
It uses 0.295M fewer repair-model tokens per task on average.
The paired median runtime reduction is 8.9 seconds.
All three comparisons use the same 72 paired tasks.

\paragraph{Qwen no-progress submission.}
\label{app:qwen-stopping}
The expanded study compares Jev-guided and unguided repair on 72
SWE-Bench Pro Verified tasks, retaining the original 36 and adding
36 distinct tasks. The judge sees five executed commands
(each output capped at 300 characters) and the next command
(400 characters). The Jev threshold is 0.31. Two consecutive low
probabilities cause a reminder; four cause submission. An unavailable
answer resets the counter.
Both conditions resolve 27 tasks. The repair-rate difference is zero, meeting the repair-loss
criterion. Jev reduces mean repair-model token use by 1.258M per task
(95\% CI: 0.905M--1.633M).
All 72 tasks have paired runtime. The paired median reduction is
655.8 seconds (95\% CI: 79.1--1,866.4).
Unguided repair reaches the 3,000-second limit in 37 runs, compared with
six under the Jev policy. The policy ends 34 runs early; unguided repair
resolves four of these tasks, and each condition resolves six tasks that
the other does not.

Same-model no-progress submission is evaluated only on the original 36-task sample,
which contains 24 historical stalled failures and 12 successes.
Its threshold is 0.0086. Jev and same-model no-progress submission each resolve
10 tasks, compared with nine without guidance. Same-model no-progress submission
reduces mean repair-model token use by 1.475M relative to unguided repair
(95\% CI: 1.016M--1.942M); its paired median runtime reduction is
972 seconds (95\% CI: 71--1,881). Jev's mean token difference from
same-model no-progress submission is $-0.182$M, and its paired median runtime
difference is $-57$ seconds.
Attributed runtime includes the inference time of the original call
when a cached answer is first reused from another run; reuse within one
run adds no inference time.

\subsection{Sampling and Estimation}

Component samples have different historical success and failure strata.
The sampling and weighting rules below define the samples represented
by their estimates; the closing paragraph relates these comparisons to
the chosen workflow.

\paragraph{Paired component estimation and sampling.}
\label{app:component-estimation}
Each DeepSeek component sample takes all failures of the original
200-task unguided run on a benchmark and an equal number of randomly
sampled successes: 36+36 on SWE-bench Verified and 59+59 on SWE-Bench
Pro Verified. Qwen's recommendation and action-type guidance samples draw equal
numbers of failures and successes
from its own unguided runs: 18+18 on SWE-Bench Pro Verified, expanded
to 59+59 for snippet recommendations with 82 added tasks, and 36+36 on
SWE-bench Verified for action-type guidance.
The no-progress and command-blocking samples initially contain
24 failures and 12 successes, expanding to 48 failures and 24 successes; the added blocking tasks reuse the unguided
controls of the no-progress extension. Stratum membership is fixed by the original 200-task unguided runs.
Unguided controls are separate runs on the same tasks, except for Qwen
snippet recommendations, which reuse the original runs. Opening estimates on SWE-Bench Pro Verified
weight the failure and success strata by their proportions in the
original 200 tasks (59/200 and 141/200 for DeepSeek, 132/200 and 68/200
for Qwen); the 72-task estimates are unweighted.
All intervals use 10,000 paired task resamples with a fixed random seed
for each analysis. Weighted opening comparisons resample within the
original success and failure strata. Weighted paired runtime medians
follow the lower inverse-CDF convention.

\paragraph{Component selection for JevCoder.}
The components selected for JevCoder meet the repair-loss criterion
with lower token use or higher observed repair rate:
snippet recommendations, action-type guidance, automatic submission on
DeepSeek, and no-progress submission on Qwen. The Gemini workflow also
uses automatic submission. Snippet recommendations replace binary file
recommendations for their higher observed repair rate on the same
DeepSeek tasks; binary file recommendations save more tokens
(Table~\ref{tab:component-intervals}). File recommendations (Score)
reduce tokens but miss the criterion, as does command blocking.
Automatic submission uses fewer repair-model tokens than submission
reminders. The stopping-signal replays
(Appendix~\ref{app:stopping-replay}; Table~\ref{tab:stopping-opportunities})
explain the model-specific
closing policies: no-progress submission rarely triggers on DeepSeek or
Gemini, whereas the Qwen successful-check signal occurs in only 57 of
400 historical runs.

\section{Full-Workflow Statistics}
\label{app:full-workflow}

Comparisons of the full JevCoder workflow test whether lower resource use
accompanies maintained repair quality on the same tasks. Each condition
has one run on 200 SWE-bench Verified and 200 SWE-Bench Pro Verified
tasks. This appendix gives the full uncertainty estimates behind
Table~\ref{tab:jevcoder-evaluation}, using the paired bootstrap procedure
in Section~\ref{sec:setup}, and the analyses that explain its runtime
and submission results. Unless stated otherwise, a difference is
condition minus the corresponding repair model's baseline; pp denotes
percentage points and M denotes one million tokens.

\paragraph{Repair, tokens, and runtime.}
Table~\ref{tab:full-intervals} reports, for each condition, the paired
repair-rate difference used to assess the prespecified repair-loss
criterion, the mean repair-model token difference, and the median of
task-level runtime differences, each with its 95\% interval. All five
repair comparisons meet the ten-point repair-loss criterion.

\begin{center}
{\fontsize{9}{11}\selectfont\setlength{\tabcolsep}{3pt}\renewcommand{\arraystretch}{1.12}%
\resizebox{\ifdim\width>\linewidth\linewidth\else\width\fi}{!}{%
\begin{tabular}{@{}lrrr@{}}
\toprule
Condition & \shortstack{$\Delta$Repair\\rate (pp)} & \shortstack{Mean\\$\Delta$Tokens (M)} & \shortstack{Paired median\\$\Delta$Runtime (s)} \\
\midrule
\multicolumn{4}{@{}l@{}}{\textit{DeepSeek-V4.1-Flash}} \\
\textbf{JevCoder} & {\boldmath$+1.00$} & {\boldmath$-0.427$} & {\boldmath$-9.5$} \\
 & {\boldmath$[-1.75,+3.75]$} & {\boldmath$[-0.632,-0.246]$} & {\boldmath$[-16.3,-3.1]$} \\
Same-model judging & $+0.75$ & $-0.272$ & $+83.2$ \\
 & $[-2.00,+3.50]$ & $[-0.446,-0.114]$ & $[+68.8,+100.1]$ \\
\midrule
\multicolumn{4}{@{}l@{}}{\textit{Qwen3.8-27B}} \\
\textbf{JevCoder} & {\boldmath$+1.75$} & {\boldmath$-0.756$} & {\boldmath$-149.3$} \\
 & {\boldmath$[-3.00,+6.50]$} & {\boldmath$[-0.931,-0.590]$} & {\boldmath$[-221.9,-87.0]$} \\
Same-model judging & $+0.25$ & $-0.532$ & $+21.7$ \\
 & $[-4.25,+4.75]$ & $[-0.703,-0.369]$ & $[-19.5,+78.6]$ \\
\midrule
\multicolumn{4}{@{}l@{}}{\textit{Gemini 3.8 Flash}} \\
\textbf{JevCoder} & {\boldmath$+2.75$} & {\boldmath$-1.456$} & {\boldmath$-123.2$} \\
 & {\boldmath$[-0.25,+5.75]$} & {\boldmath$[-1.807,-1.116]$} & {\boldmath$[-158.7,-95.0]$} \\
\bottomrule
\end{tabular}}\par}
\captionof{table}{Paired differences from the corresponding baseline on 400 tasks;
95\% intervals appear below the estimates. Token differences are per-task
means in millions and include judging requests for same-model judging;
runtime differences are medians of task-level differences and include
tools and judge inference.}
\label{tab:full-intervals}
\end{center}

\paragraph{Task-level gains and regressions.}
\label{app:task-outcomes}
The repair-rate differences above are net changes over the same tasks.
Table~\ref{tab:task-outcomes} separates each difference into tasks that
the condition newly resolves and tasks that only the baseline resolves.
JevCoder newly resolves more tasks than it loses for all three repair
models, and it keeps 95.4\% of DeepSeek's and 94.9\% of Gemini's
baseline-resolved tasks.

\begin{center}
{\fontsize{9}{11}\selectfont\setlength{\tabcolsep}{3pt}\renewcommand{\arraystretch}{1.12}%
\resizebox{\ifdim\width>\linewidth\linewidth\else\width\fi}{!}{%
\begin{tabular}{@{}lrrrrr@{}}
\toprule
Condition & Both & Gained & Regressed & Neither & Net (pp) \\
\midrule
\multicolumn{6}{@{}l@{}}{\textit{DeepSeek-V4.1-Flash}} \\
\textbf{JevCoder} & \textbf{291} & \textbf{18} & \textbf{14} & \textbf{77} & {\boldmath$+1.00$} \\
Same-model judging & 290 & 18 & 15 & 77 & $+0.75$ \\
\midrule
\multicolumn{6}{@{}l@{}}{\textit{Qwen3.8-27B}} \\
\textbf{JevCoder} & \textbf{138} & \textbf{49} & \textbf{42} & \textbf{171} & {\boldmath$+1.75$} \\
Same-model judging & 135 & 46 & 45 & 174 & $+0.25$ \\
\midrule
\multicolumn{6}{@{}l@{}}{\textit{Gemini 3.8 Flash}} \\
\textbf{JevCoder} & \textbf{259} & \textbf{25} & \textbf{14} & \textbf{102} & {\boldmath$+2.75$} \\
\bottomrule
\end{tabular}}\par}
\captionof{table}{Task-level outcomes against the corresponding baseline on
400 tasks. Both and Neither count tasks resolved by both runs or by
neither; Gained counts tasks resolved only by the condition, and
Regressed counts tasks resolved only by the baseline. Net is
100(Gained $-$ Regressed)/400, the repair-rate difference in
Table~\ref{tab:full-intervals}.}
\label{tab:task-outcomes}
\end{center}

\paragraph{Runtime decomposition.}
\label{app:runtime-decomposition}
Runtime is the sum of judge inference time and repair-model request
plus tool time. Table~\ref{tab:runtime-decomposition} reports paired median differences
separately for judge inference, repair-model and tool execution, and
total runtime. Same-model judging shortens repair-model and tool time for both models,
but its median judging time exceeds those savings. JevCoder's median
judging time is below 13 seconds for both models.

\begin{center}
{\fontsize{9}{11}\selectfont\setlength{\tabcolsep}{3pt}\renewcommand{\arraystretch}{1.12}%
\resizebox{\ifdim\width>\linewidth\linewidth\else\width\fi}{!}{%
\begin{tabular}{@{}lrrr@{}}
\toprule
Condition & Judge $\Delta$ (s) & \shortstack{Repair + tools\\$\Delta$ (s)} & Total $\Delta$ (s) \\
\midrule
\multicolumn{4}{@{}l@{}}{\textit{DeepSeek-V4.1-Flash}} \\
\textbf{JevCoder} & {\boldmath$+8.7$} & {\boldmath$-15.4$} & {\boldmath$-9.5$} \\
 & {\boldmath$[+8.1,+9.1]$} & {\boldmath$[-21.8,-9.6]$} & {\boldmath$[-16.3,-3.1]$} \\
Same-model judging & $+95.4$ & $-22.3$ & $+83.2$ \\
 & $[+91.8,+99.6]$ & $[-31.8,-11.6]$ & $[+68.8,+100.1]$ \\
\midrule
\multicolumn{4}{@{}l@{}}{\textit{Qwen3.8-27B}} \\
\textbf{JevCoder} & {\boldmath$+12.3$} & {\boldmath$-165.6$} & {\boldmath$-149.3$} \\
 & {\boldmath$[+11.3,+13.6]$} & {\boldmath$[-252.7,-100.0]$} & {\boldmath$[-221.9,-87.0]$} \\
Same-model judging & $+238.7$ & $-192.1$ & $+21.7$ \\
 & $[+223.6,+262.3]$ & $[-338.0,-84.1]$ & $[-19.5,+78.6]$ \\
\bottomrule
\end{tabular}}\par}
\captionof{table}{Paired median runtime differences from the corresponding baseline
on 400 tasks, reported separately for judge inference time and for
repair-model request plus tool time; 95\% intervals appear below the
estimates. Medians of the two parts do not add to the total. Judge time is
client time after subtracting the measured fixed connection delay; the baseline has no judge
time.}
\label{tab:runtime-decomposition}
\end{center}

\paragraph{Jev usage.}
\label{app:jev-usage}
Jev's API-reported input and output usage averages 0.152M Jev tokens per
task with DeepSeek, 0.184M with Gemini, and 0.193M with Qwen.
Each estimate covers the same 400-task sample and includes the original
usage of reused answers (Appendix~\ref{app:protocol}). Usage is measured
with Jev's tokenizer and recorded separately from repair-model tokens;
Jev tokens and repair-model tokens are not added.

\paragraph{Estimated API costs.}
\label{app:api-cost}
We estimate API costs by combining repair requests with snippet-selection
and in-repair judging, using the prices and cache charges recorded during
the experiments. Reused answers are charged at their original source
cost, so the estimate represents standalone execution of the workflow.
Table~\ref{tab:api-cost} reports 400 DeepSeek pairs and 392 Gemini pairs
with repair costs recorded in both conditions.
Including Jev's charges, JevCoder reduces estimated API cost by
4.2\% with DeepSeek and 29.7\% with Gemini; costs per resolved issue
fall by 5.4\% and 32.5\%, respectively.

\begin{center}
{\fontsize{9}{11}\selectfont\setlength{\tabcolsep}{3pt}\renewcommand{\arraystretch}{1.12}%
\resizebox{\ifdim\width>\linewidth\linewidth\else\width\fi}{!}{%
\begin{tabular}{@{}lrrr@{}}
\toprule
Condition & \shortstack{Total\\(US\$)} & \shortstack{Per task\\(US\$)} & \shortstack{Per resolved\\issue (US\$)} \\
\midrule
\multicolumn{4}{@{}l@{}}{\textit{DeepSeek-V4.1-Flash (400 paired tasks)}} \\
Unguided repair & 22.87 & 0.0572 & 0.0750 \\
\textbf{JevCoder} & \textbf{21.91} & \textbf{0.0548} & \textbf{0.0709} \\
Same-model judging & 38.53 & 0.0963 & 0.1251 \\
\midrule
\multicolumn{4}{@{}l@{}}{\textit{Gemini 3.8 Flash (392 paired tasks)}} \\
Unguided repair & 368.76 & 0.9407 & 1.3607 \\
\textbf{JevCoder} & \textbf{259.15} & \textbf{0.6611} & \textbf{0.9190} \\
\bottomrule
\end{tabular}}\par}
\captionof{table}{Estimated API costs, including repair and judging requests.
Per-task values are means. Cost per resolved issue divides costs from
all attempts by the number resolved.}
\label{tab:api-cost}
\end{center}

\paragraph{Pricing the full Gemini sample.}
For all 400 Gemini tasks, we also price the reported input and output
usage under one common scenario: US\$0.75 per million input tokens and
US\$3.75 per million output tokens, including thinking tokens, with no
discount for cached input tokens. Adding Jev's input charges gives estimated costs
of US\$4.147 per task for unguided repair and US\$3.073
for JevCoder, a 25.9\% reduction. This scenario prices reported usage.

\paragraph{Break-even prices.}
Let $S_R$ be the reduction in repair-request API cost and $U_J$ the
Jev input units per task. At a Jev price $p_J$ per million input units,
net API savings are
\[
S_R - p_J U_J / 10^6.
\]
Table~\ref{tab:list-prices} compares list prices. Jev's input price
is 4--7 times below DeepSeek-V4.1-Flash's, 18 times below Gemini~3.8
Flash's, and 48--238 times below those of widely used frontier models;
Jev output is free. The judge's cost per call is therefore small, and
the saved repair tokens are priced at the repair model's rate. For
DeepSeek, the least expensive repair model in Table~\ref{tab:api-cost},
net API savings remain positive up to a Jev price of US\$0.0585 per
million input units, 39\% above the US\$0.042 rate used in our
experiments. With Gemini, the estimated API cost falls by 29.7\% with
Jev charges included. The net advantage grows with the repair model's
price, so the same token savings would give a larger cost advantage
with the frontier models in Table~\ref{tab:list-prices}.

\begin{center}
{\fontsize{9}{11}\selectfont\setlength{\tabcolsep}{4pt}\renewcommand{\arraystretch}{1.12}%
\begin{tabular}{@{}lrr@{}}
\toprule
Model & Input (US\$/M) & Output (US\$/M) \\
\midrule
Jev 1.13 & 0.042 & 0 \\
\midrule
DeepSeek-V4.1-Flash & 0.15--0.30 & 0.60--1.20 \\
Gemini 3.8 Flash & 0.75 & 3.75 \\
\midrule
GPT-6.1 Sol & 2.00 & 10.00 \\
Claude Opus 5.5 & 4.00 & 20.00 \\
GPT-6 Astra & 10.00 & 50.00 \\
Claude Fable 5.1 & 10.00 & 50.00 \\
\bottomrule
\end{tabular}\par}
\captionof{table}{List prices per million tokens on 13 October 2026,
from the providers' pricing pages
\citep{typesafe2026,deepseekpricing2026,geminipricing2026,openaipricing2026,anthropicpricing2026}.
Jev charges input only. DeepSeek lists off-peak and peak rates for
cache misses. The top block is the judge; the middle block is the
API-served repair models in this study; the bottom block is frontier models not
tested here.}
\label{tab:list-prices}
\vspace{4pt}
\end{center}
Qwen's local GPU service is not assigned an API price.

\paragraph{Tokens per resolved issue.}
Resource use can also be expressed per unit of useful output. Dividing
repair-model tokens from all attempts by the number of resolved issues
gives 2.594M versus 2.007M for DeepSeek, 3.323M versus 1.580M for Qwen,
and 7.945M versus 5.586M for Gemini (baseline versus JevCoder).
These ratios count tokens from all attempts, including failed ones.

\paragraph{Same-model judging and submission groups.}
\label{app:gemini-submission-groups}
Holding the workflow fixed, replacing the judge changes both the
subsequent repair work and the tokens needed to obtain judgments.
Relative to same-model judging, JevCoder uses 0.155M fewer repair-model
tokens per task with DeepSeek (95\% CI: $[-0.328,+0.007]\mathrm{M}$)
and 0.225M fewer with Qwen (95\% CI: $[-0.313,-0.135]\mathrm{M}$), corresponding to
mean reductions of 9.1\% and 23.3\%, respectively.
Excluding judging requests, same-model judging reduces mean
repair-request tokens relative to the baseline by 27.2\% with DeepSeek
and 55.0\% with Qwen. Its observed mean repair-request usage is
0.110M below JevCoder's with DeepSeek and 0.066M below with Qwen;
adding judging requests reverses this ordering.
In DeepSeek's JevCoder runs, the program submits 96 of 400 attempts
directly (69 SWE-bench Verified and 27 SWE-Bench Pro Verified), and none
is followed by another repair-model request; 299 attempts end through the
repair model's own submission and five reach the time limit.
In Gemini's JevCoder runs, the program submits 133 of 400 attempts
directly, and all 133 end without another repair-model request.
JevCoder delivers a non-empty patch in 382 of 400 Gemini runs, compared
with 359 for unguided repair ($+5.75$ percentage points; 95\% CI:
$+3.00$ to $+8.75$).
Grouping tasks by whether automatic submission occurs, mean
repair-model token use is 2.027M lower than the baseline in the 133
automatically submitted attempts and 1.171M lower in the other 267.

\paragraph{Removing single components.}
\label{app:leave-one-out}
To show what each component contributes inside the complete workflow,
we remove one component at a time from the DeepSeek JevCoder workflow
on the 190-task component panel: 72 SWE-bench Verified and
118 SWE-Bench Pro Verified tasks. Each run keeps the configuration of
the same task's JevCoder run and differs only in the removed
component. Estimates are paired with that JevCoder run, use
the panel's stratum weights, and weight the two benchmarks equally;
runs without an official outcome are excluded pairwise.
Removing any single component lowers the observed repair rate and
raises mean repair-model token use and paired median runtime
(Table~\ref{tab:leave-one-out}). Runtime increases most when snippet
recommendations or automatic submission are removed.

\begin{center}
{\fontsize{9}{11}\selectfont\setlength{\tabcolsep}{3pt}\renewcommand{\arraystretch}{1.12}%
\resizebox{\ifdim\width>\linewidth\linewidth\else\width\fi}{!}{%
\begin{tabular}{@{}lrrr@{}}
\toprule
Removed component & \shortstack{$\Delta$Repair\\rate (pp)} & \shortstack{Mean\\$\Delta$Tokens (M)} & \shortstack{Paired median\\$\Delta$Runtime (s)} \\
\midrule
Snippet recommendations & $-1.49$ & $+0.174$ & $+32.3$ \\
 & $[-5.26,+1.99]$ & $[-0.047,+0.408]$ & $[+5.1,+49.3]$ \\
\addlinespace[2pt]
Action-type guidance & $-0.91$ & $+0.121$ & $+2.9$ \\
 & $[-5.92,+3.81]$ & $[-0.081,+0.326]$ & $[-10.0,+10.6]$ \\
\addlinespace[2pt]
Automatic submission & $-2.80$ & $+0.099$ & $+17.1$ \\
 & $[-7.72,+1.90]$ & $[-0.078,+0.270]$ & $[+4.7,+30.9]$ \\
\bottomrule
\end{tabular}}\par}
\captionof{table}{DeepSeek complete workflow with one component removed, relative to
the complete workflow on the same tasks (removed minus complete). 95\%
intervals appear below the estimates; 181--184 paired tasks per
comparison. Token differences are per-task means in millions.}
\label{tab:leave-one-out}
\end{center}

\end{document}